\documentclass[aps,prl,twocolumn,superscriptaddress]{revtex4-1}
\usepackage{amsmath}    % need for subequations

\usepackage{graphicx}
\usepackage{xcolor}
\definecolor{darkblue}{HTML}{3771C8}
\usepackage[%  
    colorlinks=true,
    pdfborder={0 0 0},
    citecolor = darkblue, 
    urlcolor = darkblue, 
    allcolors = darkblue
]{hyperref}
\begin{document}

%\preprint{}

\title{Frequency-Domain Quantum Interference with Correlated Photons from an Integrated Microresonator}

%\thanks{}

\author{Chaitali Joshi}

\affiliation{Applied Physics and Applied Mathematics, Columbia University, New York, NY 10027}
\affiliation{Applied and Engineering Physics, Cornell University, Ithaca, NY 14850}

\author{Alessandro Farsi}
\affiliation{Applied Physics and Applied Mathematics, Columbia University, New York, NY 10027}

\author{Avik Dutt}
\affiliation{Ginzton Laboratory and Department of Electrical Engineering, Stanford University,
Stanford, CA 94305}

\author{Bok Young Kim}
\affiliation{Applied Physics and Applied Mathematics, Columbia University, New York, NY 10027}

\author{Xingchen Ji}
\affiliation{Department of Electrical Engineering, Columbia University, New York, NY 10027}

\author{Yun Zhao}
\affiliation{Department of Electrical Engineering, Columbia University, New York, NY 10027}

\author{Andrew M. Bishop}
\affiliation{Applied Physics and Applied Mathematics, Columbia University, New York, NY 10027}

\author{Michal Lipson}
\affiliation{Applied Physics and Applied Mathematics, Columbia University, New York, NY 10027}
\affiliation{Department of Electrical Engineering, Columbia University, New York, NY 10027}

\author{Alexander L. Gaeta}
\email[Corresponding Author:]{a.gaeta@columbia.edu }
\affiliation{Applied Physics and Applied Mathematics, Columbia University, New York, NY 10027}
\affiliation{Department of Electrical Engineering, Columbia University, New York, NY 10027}


%\date{\today}
% In the abstract I am going to sell two things- frequency and integrated and high visibility. Say something like microring resonators enable scaling and sell nonlinear optics, control light with light
%Read anton zeilingers OAM papers to find similar selling points for frequency qubits. 
%Question: can you claim that 

\begin{abstract}
Frequency encoding of quantum information together with fiber and integrated photonic technologies can significantly reduce the complexity and resource requirements for realizing all-photonic quantum networks. The key challenge for such frequency domain processing of single photons is to realize coherent and selective interactions between quantum optical fields of different frequencies over a range of bandwidths. Here, we report frequency-domain Hong-Ou-Mandel interference with spectrally distinct photons generated from a chip-based microresonator. We use four-wave mixing to implement an active `frequency beam-splitter' and achieve interference visibilities of $0.95 \pm 0.02$. Our work establishes four-wave mixing as a tool for selective high-fidelity two-photon operations in the frequency domain which, combined with integrated single-photon sources, provides a building block for frequency-multiplexed photonic quantum networks. 
%We report quantum interference with spectrally distinct correlated photons generated from an integrated microring resonator. We implement a tunable frequency-beam splitter using Bragg-Scattering Four-Wave Mixing (BS-FWM). We achieve a background limited interference visibility of $95\pm0.02$. Our results combine BS-FWM together with integrated single-photon sources to enable the generation of multi-photon frequency-entangled cluster states on an integrated photonics platform. 
\end{abstract}
% In the abstract I must precisely sel fourwave mixing, integrated single photon sources, the concept of a tunable frequency beam splitter, the high visibility 

%\keywords{}

\maketitle

%\section{Introduction}
Two-photon interference is a fundamental quantum effect with no classical analogue. Such interference is at the heart of photonic quantum information processing (QIP) and is the basis of several QIP realizations such as Bell-state measurement, boson sampling, measurement-based logic gates, and the generation of multipartite entangled Greenberger-Horne-Zeilinger (GHZ) and cluster states \cite{bouwmeester1999, knill2001, spring2013, ewert2014}. In the original Hong-Ou-Mandel (HOM) interference experiment, the photon wave packets are in distinct spatial modes \cite{hong1987a}, but are otherwise identical in their polarization, spectral and temporal properties at the input. However, subsequent experimental and theoretical work confirms it is the indistinguishability of the two-photon amplitudes at the output of the interferometer that is crucial to the observation of HOM-type interference \cite{pittman1996}. This leads to the interesting possibility of observing quantum interference involving spectrally distinct photons. An `active' device that coherently mixes two input frequency modes can render distinct spectral amplitudes indistinguishable, resulting in fourth-order interference with the two photons bunched in the same frequency mode \cite{raymer2010}. 

%Pasting here deleted text for result.At first, this is counterintuitive, as such spectrally distinct two-photon amplitudes do not interfere on passive devices such as spatial or polarization mode beam splitters. However, when a single photon is incident on one port of such a frequency beam splitter, the resulting output is a photon in a superposition of two distinct frequencies; a bichromatic qubit \cite{clemmen2016}. When two photons of distinct frequencies are incident on different ports of this device, the two-photon amplitude where both photons undergo frequency conversion destructively interferes with the two-photon amplitude where neither photon gets frequency converted. This results in Hong-Ou-Mandel type interference, with both photons bunching in the same frequency mode with identical frequencies. 

In this work, we combine frequency-entangled photons generated on-chip together with Bragg-scattering four-wave mixing (BS-FWM) in optical fiber to demonstrate frequency-domain HOM interference visibilities as high as 95\%. We show interference with narrow photons less than 300 MHz in bandwidth and widely separated in frequency by 800 GHz. We predict and observe a rich two-photon interference pattern, including the phenomenon of quantum beating in the temporal domain. Novel approaches for all-photonic quantum repeaters rely on a combination of efficient single-photon sources, linear operations with beam splitters and measurement-based fusion gates to generate entangled multi-photon Greenberger-Horne-Zeilinger (GHZ) and cluster states \cite{zwerger2012a, azuma2015}. Frequency multiplexing can massively reduce the resource requirements for such all-photonic quantum networks. Frequency domain quantum operations also provide a distinct advantage in terms of the scaling of losses over spatial or polarization mode processing \cite{lukens2017, joshi2018a, hiemstra2019}. Cavity-enhanced spontaneous four-wave mixing (SFWM) in integrated microresonators produces frequency-entangled photon pairs with an effectively discrete joint spectral intensity \cite{reimer2014, harris2014, ramelow2015, jiang2015, grassani2015}. A large number of such compact, identical sources can be integrated on a monolithic platform for the generation and manipulation of complex non-classical states of light \cite{montaut2017, meyer-scott2018, qiang2018, paesani2019}. Our demonstration establishes BS-FWM as a tool for selective, high-fidelity two-photon operations between frequency modes of integrated microresonators. %we demonstrate two-photon quantum interference with up tp 95\% visibility by combining spectrally distinct photons generated on a silicon nitride chip together with Bragg scattering four-wave mixing (BS-FWM) in a dispersion-shifted telecom fiber\cite{clemmen2016}. Our work aims to comb

\begin{figure}[t]
\includegraphics[width=0.5\textwidth]{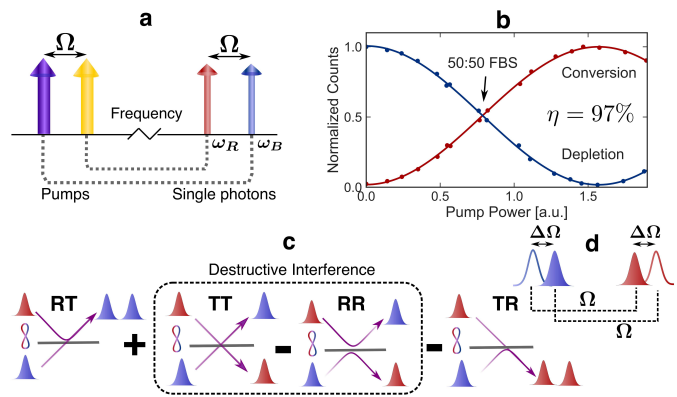}
\caption{Frequency beam splitter via BS-FWM: \textbf{a)} Two strong classical pumps $(\omega_{P1}, \omega_{P2})$ mediate the interaction between two quantum fields, $(\omega_R, \omega_B)$ in a medium with third-order $\chi^{(3)}$ nonlinearity. \textbf{b)} Measured signal conversion and depletion (efficiency $\eta = 97\%$). BS-FWM acts as a 50:50 frequency beam splitter (FBS) when the pump power is such that the nonlinear interaction strength $\gamma PL = \pi/8$. \textbf{c)} Two frequency-correlated photons incident on a 50:50 FBS. Perfect HOM-type interference is observed when the two-photon amplitude for both photons being frequency translated (RR) and neither photon being translated (TT) are indistinguishable. This occurs when the BS-FWM pump separation $\Omega$ matches the input photon-pair separation, resulting in $\Delta\Omega = 0$ (see \textbf{d}). \label{fig:blochsphere}} 
%When the pump power is such that $2\gammaPL = \frac{pi}{4}$, BS-FWM acts as a frequency beam-splitter and corresponds to a  $\ffrac{\pi}{2}$ rotation on the Bloch sphere.}
\end{figure}

%Creating a device that coherently combines distinct frequency modes is challenging as this requires a strong noise-free nonlinear process. 
 BS-FWM is a unitary, third-order parametric process in which two strong classical pump waves mediate the interaction between the quantum fields via a third order ($\chi^{(3)}$) nonlinearity (Figs. \ref{fig:blochsphere}a and b)\cite{mcguinness2010, farsi2015a, mckinstrie2004,mckinstrie2005a, mckinstrie2005}. By controlling the power and phase of the classical fields involved, we emulate a tunable `active' frequency beam splitter \cite{clemmen2016}. For quantum frequency conversion based on $\chi^{(2)}$ nonlinearity, the input and target modes must be placed in different optical bands, typically separated by few tens to a hundred terahertz in order to satisfy energy conservation. Alternatively, electro-optic modulators (EOMs) can impart only small frequency shifts of the order of a few gigahertz. With BS-FWM, the use of two classical fields provides an additional degree of freedom and efficient conversion is possible for separations ranging from a few hundred GHz to a few THz \cite{joshi2018a}. This makes a BS-FWM frequency beam splitter (FBS) compatible with the typical free spectral range (FSR) of integrated microresonators and also with dense wavelength division multiplexing (DWDM) components aligned to the ITU grid. Previously, EOMs have been used in combination with integrated sources to generate high-dimensional time-frequency entangled states  \cite{kues2017, reimer2019, imany2019}. Together with pulse shaping and bulk single-photon sources, EOMs have also been used to create frequency-bin entangled states and to demonstrate two-photon HOM-type interference with modes separated by up to 25 GHz \cite{merolla1999a, olislager2010,  olislager2014, lu2018, imany2018, lu2018a, galmes2019a}. Recently, frequency translation in a $\chi^{(2)}$ crystal was used to demonstrate interference between a single photon and attenuated coherent laser light \cite{kobayashi2016}. 

% A photon in the input frequency mode $\omega_R$ is annihilated for every photon generated in the target mode $\omega_B$, and vice versa, resulting in a unitary process

%BS-FWM can then be used to selectively manipulate photon correlations by precisely matching the frequency separation of the pumps to that of the target resonator modes.
 
We use quantum frequency translation via BS-FWM to create a coherent interaction between two quantum fields at different frequencies $\left(\omega_R, \omega_B \right)$, as shown in Fig. \ref{fig:blochsphere}. Energy conservation requires that the frequency separation between the classical pump fields ($\omega_{P1} - \omega_{P2} = \Omega$) should match the separation of the quantum fields ($\omega_{B} - \omega_{R} = \Omega$). Phase-matching ($\Delta\beta = \beta_{R} + \beta_{P1} - \beta_B - \beta_{P2}$, $\beta$: propagation constant) can be ensured by placing the pump fields and the quantum fields symmetrically about the zero group-velocity dispersion (GVD, $\beta^{(2)} = 0$) point of the interaction medium. Due to the effects of third-order dispersion ($\beta^{(3)}$), it is possible to ensure selective phase matching such that the resulting process is a two-mode interaction without spurious side-bands (see supplementary Section S1). For such a selectively phased-matched process ($\Delta\beta =0$), the mode transformations for signal and idler annihilation operators $\hat{a}(\omega_R), \hat{a}(\omega_B)$ after passing through the FBS are given by, 
%For a given pump separation $\Omega$ and an input quantum field $\omega$, energy conservation is satisfied for two translated sidebands $\omega - \Omega$ and $\omega + \Omega$. 
\begin{align}
&\hat{a}_R(\omega_R)  \rightarrow \upsilon \hat{a}_R(\omega_R) - \mu \hat{a}_B(\omega_R + \Omega),   \nonumber \\ 
&\hat{a}_B(\omega_B) \rightarrow  \mu^{\ast}  \hat{a}_R(\omega_B - \Omega) + \upsilon^{\ast} \hat{a}_B(\omega_B),   \label{eq:FBS}
\end{align} 
where $\upsilon = \cos {(2\gamma PL)}, \mu = e^{i\phi}\sin{(2\gamma P L)}$,  $\Omega$ is the frequency separation of the two pump fields, $P = \sqrt{P_{1} P_{2}}$ depends on the power $P_1, P_2$ of the two classical pumps, $\gamma$ is proportional to the nonlinearity of the interaction medium, $L$ is the interaction length, $\phi$ is the relative phase between the pumps. The subscripts $(R,B)$ denote the red-detuned and blue-detuned frequencies respectively. As evident from Eq. \ref{eq:FBS} and as demonstrated in Ref. \cite{clemmen2016}, BS-FWM acts as an ${SU}(2)$ transformation and can produce arbitrary single-qubit rotations in the two-dimensional frequency Hilbert space $\{|\omega_R\rangle, |\omega_B\rangle\}$. 
%By controlling the pump power, interaction length and the relative phase of the classical pumps fields, the effective transmission ($|\tau|^2  = \cos^{2} \left(2\gamma P L\right)$) and reflection ($|\rho|^2 = \sin^{2} \left(2\gamma P L\right)$) coefficients of this beam-splitter can be varied.

%In particular when the pump parameters are set such that $\gamma P L =\pi/8$, the result is a 50:50 frequency beam splitter 
Here, we formally establish that this process can be used for two-photon quantum interference (see Figs. \ref{fig:blochsphere}b and c and Supplementary Section 2). We theoretically predict the conditions for observing perfect Hong-Ou-Mandel type interference for two frequency-correlated photons and show that our experimental findings are in excellent agreement with these theoretical predictions. The initial two-photon wave-function is given as, 
\begin{align}
|\psi\rangle = \int d\omega_{B} d\omega_{R} \: \phi(\omega_B, \omega_R)\hat{a}_R^{\dagger}(\omega_R) \hat{a}_B^{\dagger}(\omega_B)|0,0\rangle,  \label{eq:JSA}
\end{align}
where $\phi(\omega_R,\omega_B)$ is the joint spectral amplitude (JSA) of the two photons. The coincidence measurement at the output corresponds to the two-photon correlation function, 
\begin{align}
G^{(2)}(\tau) = \langle E_{R}^{(-)}(t)E_{B}^{(-)}(t+\tau)E_{B}^{(+)}(t+\tau)E_{R}^{(+)}(t)\rangle.  \label{eq:g2}
\end{align} 
The quantized electric field is related to the annihilation operator as $E^{(+)}(t) = [E^{(-)}(t)]^\dagger  \propto 1/\sqrt(2\pi) \int d\omega  \: \hat{a}(\omega)e^{-i \omega t}$. Combining Eqs. \ref{eq:JSA} and \ref{eq:g2} and using standard commutation relations we obtain,
\begin{align}
\begin{split}
%G^{(2)}(\tau) = \int  d\omega_R^{'} d\omega_B^{'} d\omega_R d\omega_B \: & e^{i(\omega_R^{'} - \omega_R)(t+ \tau)} e^{i(\omega_B^{'} - \omega_B)t} \\ & \phi^\ast(\omega_B^{'}, \omega_{R}^{'}) \phi(\omega_B, \omega_{R})     \label{eq:g2step1}\\ 
G^{(2)}(\tau) = \int  d\omega_R^{'} d\omega_B^{'} d\omega_R d\omega_B \: & e^{i(\omega_R^{'} - \omega_R)( \tau)} \phi^\ast(\omega_B^{'}, \omega_{R}^{'}) \phi(\omega_B, \omega_{R}).     \label{eq:g2step1}\\ 
\end{split}
\end{align}
As expected, the second-order correlation function is the Fourier transform of the joint spectral intensity (JSI) of the photons. We calculate the second-order correlation function after the photons have undergone the transformation in Eq. \ref{eq:FBS}. Combining Eqs. \ref{eq:FBS}, \ref{eq:JSA} and  \ref{eq:g2} for the case $\upsilon = \mu = 1/\sqrt2 $, we obtain,

%\begin{equation}
%\setlength{\jot}{2pt}
%\begin{split}
%G^{(2)}(\tau) = \int  &d\omega_R^{'} d\omega_B^{'} d\omega_R d\omega_B \:  e^{i(\omega_R^{'} - \omega_R)(t+ \tau)} e^{i(\omega_B^{'} - \omega_B)t} \\
   %& \:\:\:\:  [|\tau|^4  \phi^{\ast}(\omega_B^{'}, \omega_R^{'})\phi(\omega_B, \omega_R)  \\
                                                                                                 %& + |\rho|^4 \phi^{*}(\omega_R^{'} + \Omega, \omega_B^{'} - \Omega)\phi(\omega_R + \Omega, \omega_B -\Omega) \\  
                                                                                                 %& - |\tau|^2 |\rho|^{2}    \phi^{\ast}(\omega_B^{'}, \omega_R^{'})\phi(\omega_R + \Omega, \omega_B - \Omega) \\
                                                                                                 %& - |\tau|^2 |\rho|^{2}   \phi^{\ast}(\omega_R^{'} + \Omega, \omega_B^{'} - \Omega) \phi(\omega_B, \omega_R)]   \label{eq:g2BS}
%\end{split}
%\end{equation}
%.   \:\:\:\:\:\:

\begin{widetext}
\begin{equation}
\setlength{\jot}{2pt}
\begin{split}
G^{(2)}(\tau) = \frac{1}{4}& \int  d\omega_R^{'} d\omega_B^{'} d\omega_R d\omega_B \:  e^{i(\omega_R^{'} - \omega_R)(\tau)}  \:\:\:\: [ \phi^{\ast}(\omega_B^{'}, \omega_R^{'})\phi(\omega_B, \omega_R) - \phi^{\ast}(\omega_B^{'}, \omega_R^{'})\phi(\omega_R + \Omega, \omega_B - \Omega)  \\
& + \phi^{\ast}( \omega_R^{'} + \Omega, \omega_B^{'} - \Omega)\phi(\omega_R + \Omega, \omega_B -\Omega) -  \phi^{\ast}(\omega_R^{'} + \Omega, \omega_B^{'} - \Omega) \phi(\omega_B, \omega_R)],   
                                                                                               \label{eq:g2BS}
\end{split} 
\end{equation}
\end{widetext}
where we assume that the integration limits are well within the phase-matching bandwidth of BS-FWM. From Eq. \ref{eq:g2BS}, it is seen that perfect destructive interference occurs when the two-photon amplitude associated with both photons being frequency translated is indistinguishable from the two-photon amplitude for the case in which neither photon undergoes translation, that is when $\phi(\omega_B,\omega_R) = \phi(\omega_R + \Omega, \omega_B - \Omega)$. %This result can be interpreted as follows: if  the corresponding probability amplitudes destructively interfere resulting in Hong-Ou-Mandel type interference. 

Eqs. \ref{eq:g2step1} and \ref{eq:g2BS} can be explicitly evaluated for the case of photon pairs generated via CW-pumped SFWM in a ring resonator. The JSA for photon pairs generated in two symmetrically placed resonances $(\omega_R^0, \omega_B^0)$ about the resonance corresponding to the SFWM pump is, 
$\phi(\omega_B, \omega_R) \propto \delta(\omega_R + \omega_B - 2\omega_P) l(\omega_B, \omega_B^0)l(\omega_R, \omega_R^0)$ where $\omega_P$ is the SFWM pump frequency, $l(\omega, \omega^0) = {\left(\frac{\Delta\omega}{2}\right)^{\frac{1}{2}}}/{[-i(\omega - \omega^0) + \frac{\Delta\omega}{2}]}$, such that $|l(\omega,\omega^0)|^2$ describes the Lorentzian response of a ring resonance centered at $\omega^0$ with a full-width at half maximum $\Delta\omega$ \cite{vernon2017}. 
%We assume that the SFWM pump frequency is detuned by $\Delta$ w.r.t the ring resonance, so that $\omega_B^0 + \omega_R^0 = 2(\omega_P + \Delta)$. 
Equation \ref{eq:g2step1} then reduces to, 
\begin{align}
%G^{(2)}(\tau)  \propto \frac{\left(\frac{\Delta\omega}{2}\right)^{2}}{2\pi \left(\Delta^2 +\left(\frac{\Delta\omega}{2}\right)^{2}\right)} e^{-\Delta\omega|\tau|}
G^{(2)}(\tau)  \propto e^{-\Delta\omega|\tau|}. 
\end{align}
Similarly, evaluating Eq. \ref{eq:g2BS} results in the second-order coherence function after the 50:50 FBS, 
%G^{(2)}(\tau) =  & \propto \frac{\left(\frac{\Delta\omega}{2}\right)^{2}}{2\pi \left(\Delta^2 +\left(\frac{\Delta\omega}{2}\right)^{2}\right)} e^{-\Delta\omega |\tau|}  \\ 
%&\left(|\tau|^4 + |\rho|^4 -2|\rho|^{2}|\tau|^2 e^{-i\Delta\Omega \tau} \cos{(\Delta\Omega\tau)} \right) \\
%\begin{align}
%\begin{split}t
%G^{(2)}(\tau)  \propto  e^{-\Delta\omega |\tau|} \left(|\tau|^4 + |\rho|^4 -2|\rho|^{2}|\tau|^2 \cos{(\Delta\Omega\tau)} \right) 
%\end{split}
%\end{align}
%Setting $\rho = \dfrac{1}{\sqrt{2}}, \tau = \dfrac{1}{\sqrt{2}}$ results in further simplification of the above expression:
%G^{(2)}(\tau) & \propto   \frac{1}{2\pi (4\Delta^2 + \Delta\omega^2)} e^{-\Delta\omega |\tau|} \sin^{2}{\left(\frac{\Delta\Omega\tau}{2}\right)} \label{eq:beating}
\begin{align}
\begin{split}
G^{(2)}(\tau)  \propto e^{-\Delta\omega |\tau|} \sin^{2}{\left(\frac{\Delta\Omega\tau}{2}\right)}, \label{eq:beating}
\end{split}
\end{align}
where we have introduced the variable $\Delta\Omega = \Omega - (\omega_B^0 - \omega_R^0)$ to reflect the offset between the frequency separation $\Omega$ of the BS-FWM pumps with respect to the separation of the photon envelopes centered at $\omega_B^0$ and $\omega_R^0$. Equation \ref{eq:beating} shows that the observed interference depends sensitively on the offset $\Delta\Omega$. When $\Delta\Omega = 0$, $G^{(2)}(\tau) = 0$ for all $\tau$, as expected from the fact that two-photon wavefunction before and after frequency translation are completely indistinguishable. For offsets $\Delta\Omega$ that are small or comparable to the resonator linewidth $\Delta\omega$, the phenomenon of quantum beating is predicted \cite{legero2004b}. We note that this beating can only be resolved if $2\pi/\Delta\Omega$ is much larger than the temporal resolution of the single-photon detectors \cite{legero2003}. 

\begin{figure*}[t]
\includegraphics[width=0.75\textwidth]{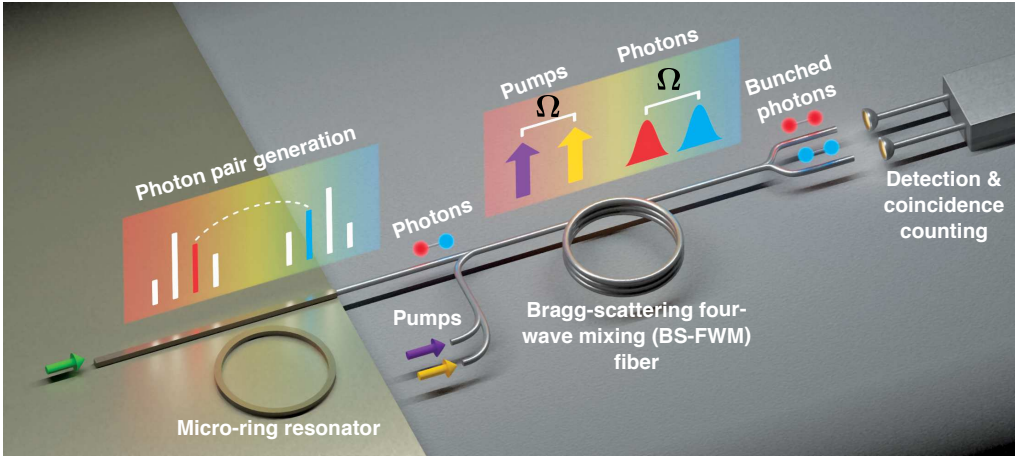}
\caption{\textbf{Frequency domain two-photon interference via BS-FWM}: Narrowband frequency-correlated photons are generated via spontaneous four-wave mixing (SFWM) in a silicon nitride microring resonator. The generated photons are coupled together with two classical pump waves through a WDM into a dispersion-shifted fiber for Bragg-scattering four-wave mixing (BS-FWM). The power of the BS-FWM pumps is set such that it acts as a 50:50 frequency beam splitter and their frequency separation $\Omega$ is set to precisely match the selected pair of correlated photons. After the nonlinear interaction, the two frequency arms are separated through a WDM followed by detection and coincidence counting. Hong-Ou-Mandel type interference is observed and both photons are bunched in the same frequency mode. \label{fig:resonator_BS_FWM}} 
%When the pump power is such that $2\gammaPL = \frac{pi}{4}$, BS-FWM acts as a frequency beam-splitter and corresponds to a  $\ffrac{\pi}{2}$ rotation on the Bloch sphere.}
\end{figure*}

We experimentally demonstrate this detuning-dependent HOM interference in agreement with these theoretical predictions using the scheme depicted in Fig. \ref{fig:resonator_BS_FWM}. A silicon nitride microresonator is pumped with a continuous-wave (CW) laser at 1282.8 nm to generate frequency-correlated photon pairs in the O-band through SFWM. The generated photons are coupled into a single-mode fiber and combined with the classical pump fields located in the C-band for BS-FWM and sent to a dispersion-shifted fiber (Corning Vistacor). The BS-FWM pumps are intensity modulated to generate 10-ns long pulses that are amplified with an erbium-doped fiber amplifier. The polarizations of the BS-FWM pumps are aligned to the linear polarization of the input state to achieve more than 95\% depletion. The pump power is then set to obtain 50\% depletion such that the photons now see a device that emulates a 50:50 FBS. The two frequency arms are separated using a WDM followed by free space grating filters to extract photons that are red detuned (centered at $\omega_R^0$) and blue detuned (centered at $\omega_B^0$) by two FSRs with respect to the SFWM pump.  The photons are then detected with superconducting nanowire single photon detectors, followed by coincidence counting using a time-tagging module. In order to match the BS-FWM pump separation $\Omega$ precisely to that of the two photons ($\omega_B^0 - \omega_R^0$), we perform a precise measurement of the FSR with a phase modulator (see Supplementary Section 3). We measure an FSR of 201.275 GHz, resulting in a photon separation ${(\omega_B^0 - \omega_R^0)}/{2\pi}$ = 805.1 GHz. A detailed description of the experimental system, including characterization of losses is included in Supplementary Section 3. 

\begin{figure*}[t]
\includegraphics[width=1\textwidth]{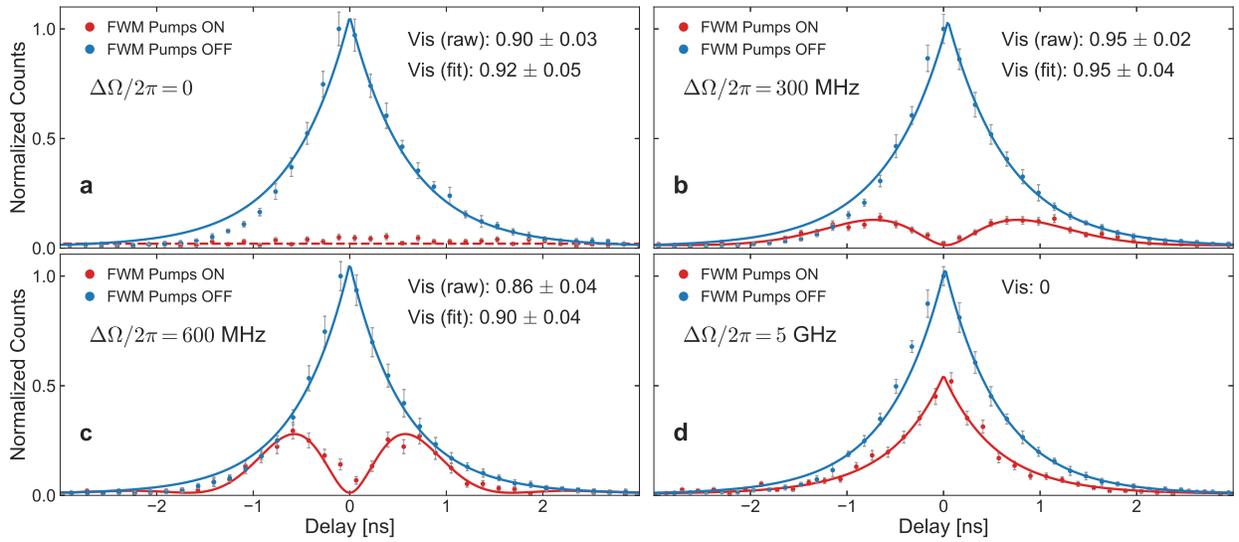}
\caption{\textbf{Experimental observation of frequency domain two-photon interference}. Red curves and blue curves (dots: experiment, solid line: fit) are the normalized three-fold coincidence counts when the BS-FWM pumps are on and off, respectively. The photon bandwidth is measured to be 270 $\pm$ 15 MHz. \textbf{a)} When $\Delta\Omega = 0$, we observe a flat $G^{(2)}(\tau)$ with a raw visibility $\alpha_r = 0.90 \pm 0.03$ (fit visibility $\alpha_f = 0.92 \pm 0.05$. \textbf{b)} For a detuning $\Delta\Omega/2\pi =300$ MHz, we observe temporal beating with a raw visibility $\alpha_r = 0.95 \pm 0.02$ (fit visibility $\alpha_f = 0.95 \pm 0.04$). \textbf{c)} For $\Delta\Omega/2\pi$ = 600 MHz, we observe an increase in the amplitude of the side lobes of the beating signal with a measured raw visibility of $\alpha_r = 0.86 \pm 0.04$ (fit visibility $\alpha_f = 0.90 \pm 0.04$). \textbf{d)} For very large detuning ($\Delta\Omega/2\pi$ = 5 GHz), the interference fringes are no longer resolved resulting in a double-exponential output ($\alpha = 0)$. Error bars are calculated assuming Poisson statistics. \label{fig:HOM}}
\end{figure*}

Our experimental results are shown in Fig. \ref{fig:HOM} and are in excellent agreement with our theoretical predictions. We obtain a photon bandwidth ${\Delta\omega}/{2\pi} = 270 \pm 15$ MHz  from the measured cross-correlation $G^{(2)}(\tau)$. In order to post-select only those photon coincidences that occur within the 10-ns temporal window of the BS-FWM pumps, we perform a three-fold coincidence measurement with the arrival time of the two photons at the SNSPDs and a synchronization signal from the pumps. We then obtain suitable normalization by averaging over coincidences accumulated in 10-ns temporal windows that are not synchronized with the BS-FWM pumps (Fig. \ref{fig:HOM}, blue curves, see also Supplementary Section 3). For an integration time of one hour, we measure 2700 three-fold normalization coincidence counts within the $1/e$ coherence time of the photons. We introduce a visibility parameter $\alpha$ in Eq. \ref{eq:beating}, such that $G^{(2)}(\tau) \propto e^{-\Delta\omega |\tau|} {\left(\frac{1}{2} - \frac{\alpha}{2}\cos{\Delta\Omega\tau}\right)}$. This visibility parameter $\alpha$ corresponds the depth of the beating signal at $\tau = 0$ and is a direct indicator of the fidelity of our 50:50 FBS and the indistinguishability of the two-photon amplitude before and after frequency translation. Any distinguishability in other degrees of freedom such as polarization or deviation from the balanced splitting ratio degrades this extinction. As expected, when the offset $\Delta\Omega$ is zero, we observe nearly complete destructive interference resulting in a flat $G^{(2)}(\tau)$. We measure a raw visibility of $\alpha_r = 0.90 \pm 0.03$ from the data and extract a visibility $\alpha_f = 0.92 \pm 0.05$ from fit to the data (Fig. \ref{fig:HOM}a, red curve). For ${\Delta\Omega}/{2\pi} =300$ MHz, we obtain a raw visibility of $\alpha_r = 0.95 \pm 0.02$ (fit visibility $\alpha_f = 0.95 \pm 0.04$, Fig. \ref{fig:HOM}b). The high visibility of these measurements indicates that BS-FWM preserves the polarization and spatio-temporal modes of the input quantum fields after frequency translation. For higher detunings $\Delta\Omega$ = 600 MHz, we see the expected increase in the amplitude of the side lobes (Fig. \ref{fig:HOM}c). We measure an interference visibility of $\alpha_r = 0.86 \pm 0.04$ (fit visibility $\alpha_f = 0.90 \pm 0.04$) from this detuning. The reduced raw visibilities in these measurements are due to fluctuations in polarization and pump power during the hour-long measurement, and due to multi-photon noise (see Supplementary Section 3) \cite{christ2012, bonneau2015}. For large detuning, $\Delta\Omega/{2\pi} = 5$ GHz, the interference fringes are no longer resolved by the detection system, resulting in a double-exponential output ($\alpha = 0$) as shown in Fig. \ref{fig:HOM}d. Our results are summarized in Table 1. The measured raw visibilities are consistent with the visibilities extracted from fit to the data within our measurement error, indicating excellent agreement between our theoretical predictions and experiment. 
\begin{table}
\begin{tabular}{|c|c|c|} 
\hline
Detuning $\Delta\Omega$ (MHz) &  Raw Visibility ($\alpha_r) $ & Fit Visibility ($\alpha_f$)  \\ \hline
0 & 0.90 $\pm$ 0.03 & 0.92 $\pm$ 0.05 \\
300 & 0.95 $\pm$ 0.02 & 0.95 $\pm$ 0.04 \\
600 & 0.86 $\pm$ 0.04 & 0.90 $\pm$ 0.04 \\ \hline
\end{tabular}
\caption{\textbf{Measured two-photon interference visibilities:} The measured raw visibilities ($\alpha_r$) are consistent with the visibilities extracted from fit to the data ($\alpha_f$) within our measurement error, indicating excellent agreement between our theory and experiment. \vspace{-13pt}}
\end{table}

%We emphasize that we are observing interference of the joint two-photon amplitudes before and after the nonlinear process, rather than that of the individual probability amplitudes of the photons themselves. This is evident from the fact that the photons travel a distance that is much longer than the coherence length of the SFWM pumps in the 100-m long BS-FWM fiber \cite{pittman1996}. 

%Say here that it is the two-photon amplitude that matters and not the individual amplitude of each photon

%\begin{align}
%\begin{split}
%G^{(2)}(\tau) & \propto \frac{1}{2\pi (4\Delta^2 + \Delta\omega^2)} e^{-\Delta\omega |\tau|} {\left(\frac{1}{2} - \frac{\alpha}{2}\cos\frac{\Delta\Omega\tau}{2}\right)} \label{eq:beating}
%\end{split}
%\end{align}

\begin{figure}[t]
\centering
\includegraphics[width=0.45\textwidth]{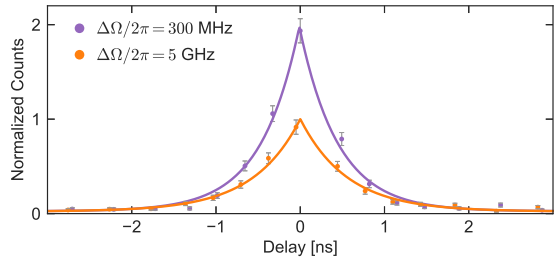}
\caption{\textbf{Photon bunching via autocorrelation}. Measured enhancement in two-photon bunching for small (purple, $\Delta\Omega/2\pi = 300$ MHz) over large pump detunings (orange, $\Delta\Omega/2\pi = 5$ GHz) with a measured peak autocorrelation  of $1.93 \pm 0.13$. \vspace{-10pt} \label{fig:G2}}
%When the pump power is such that $2\gammaPL = \frac{pi}{4}$, BS-FWM acts as a frequency beam-splitter and corresponds to a  $\ffrac{\pi}{2}$ rotation on the Bloch sphere.}
\end{figure}

The 50:50 FBS stochastically bunches photons to the same frequency mode with a probability ${1}/{4}$ even in the absence of two-photon interference (Fig. \ref{fig:HOM}d, red curve). This probability is enhanced by a factor of 2 in the presence of two-photon interference, resulting in near-perfect coalescence. We experimentally confirm this enhancement in bunching with a second-order auto correlation measurement on the blue-detuned ($\omega_B$) frequency arm. As shown in Fig. \ref{fig:G2}, we observe an enhancement for small pump detunings ${\Delta\Omega}/{2\pi}$ = 300 MHz as compared to the case with large pump detunings ${\Delta\Omega}/{2\pi} = 5$ GHz, with a measured peak autocorrelation of $1.93 \pm 0.13$.

%In conclusion, our work generalizes BS-FWM from a technique for frequency translation to a highly versatile toolbox for coherent and selective manipulation of quantum fields in the frequency domain. We observe quantum interference in a single spatial mode %of a fiber, and therefore do not require active stabilization or alignment of interferometric paths to ensure stability. In addition to fundamental novelty, our work can significantly reduce the losses and complexity of experiments such boson sampling and %multipartite-entangled GHZ state generation. With conventional spatial and polarization mode interferometers, the depth of the circuit and photon loss increase as more modes are included for processing. With BS-FWM, complex frequency-domain operations %are possible by including auxiliary pumps in the nonlinear interaction without adding loss to the path of the photons. Our demonstration in combination with integrated single-photon sources creates a promising path for frequency-encoded photonic quantum %networks. 
In conclusion, we have demonstrated two-photon interference in the frequency domain using an on-chip photon source with visibilities as high as 95\%. In contrast to experiments based on bulk photon sources and free-space optics, we observe interference in a single spatial mode, and do not require active stabilization or alignment of interferometric paths to achieve high visibility. Selective two-photon operations are possible between arbitrary pairs of resonator modes over a large bandwidth up to a few THz (see Supplementary Figure 1d) \cite{joshi2018a}. While multiplexing several two-photon operations will be associated with a classical resource overhead for the preparation of additional BS-FWM pumps, no additional components or losses are introduced in the paths of the photons (see Supplementary Section 4). Our demonstration can be extended to spectrally pure single-photons generated via pulsed excitation of the microresonator \cite{vernon2017, roztocki2017}. Together with implementations of BS-FWM in nanophotonic devices \cite{agha2012,li2016a, li2017,bell2017}, such two photon operations can be used for the on chip generation of multipartite entangled GHZ and cluster states. Our demonstration offers a path to combining on-chip photon sources, fiber-based wavelength division multiplexing and four-wave mixing for the realization of scalable frequency-multiplexed photonic quantum repeaters and networks. 

\newpage
\subsection*{Acknowledgements}
We acknowledge useful discussions with Aseema Mohanty, Yoshitomo Okawachi, Rajveer Nehra, Ben Sparkes and Jakob Gillespie. 

\begin{thebibliography}{48}%
\makeatletter
\providecommand \@ifxundefined [1]{%
 \@ifx{#1\undefined}
}%
\providecommand \@ifnum [1]{%
 \ifnum #1\expandafter \@firstoftwo
 \else \expandafter \@secondoftwo
 \fi
}%
\providecommand \@ifx [1]{%
 \ifx #1\expandafter \@firstoftwo
 \else \expandafter \@secondoftwo
 \fi
}%
\providecommand \natexlab [1]{#1}%
\providecommand \enquote  [1]{``#1''}%
\providecommand \bibnamefont  [1]{#1}%
\providecommand \bibfnamefont [1]{#1}%
\providecommand \citenamefont [1]{#1}%
\providecommand \href@noop [0]{\@secondoftwo}%
\providecommand \href [0]{\begingroup \@sanitize@url \@href}%
\providecommand \@href[1]{\@@startlink{#1}\@@href}%
\providecommand \@@href[1]{\endgroup#1\@@endlink}%
\providecommand \@sanitize@url [0]{\catcode `\\12\catcode `\$12\catcode
  `\&12\catcode `\#12\catcode `\^12\catcode `\_12\catcode `\%12\relax}%
\providecommand \@@startlink[1]{}%
\providecommand \@@endlink[0]{}%
\providecommand \url  [0]{\begingroup\@sanitize@url \@url }%
\providecommand \@url [1]{\endgroup\@href {#1}{\urlprefix }}%
\providecommand \urlprefix  [0]{URL }%
\providecommand \Eprint [0]{\href }%
\providecommand \doibase [0]{https://doi.org/}%
\providecommand \selectlanguage [0]{\@gobble}%
\providecommand \bibinfo  [0]{\@secondoftwo}%
\providecommand \bibfield  [0]{\@secondoftwo}%
\providecommand \translation [1]{[#1]}%
\providecommand \BibitemOpen [0]{}%
\providecommand \bibitemStop [0]{}%
\providecommand \bibitemNoStop [0]{.\EOS\space}%
\providecommand \EOS [0]{\spacefactor3000\relax}%
\providecommand \BibitemShut  [1]{\csname bibitem#1\endcsname}%
\let\auto@bib@innerbib\@empty
%</preamble>
\bibitem [{\citenamefont {Bouwmeester}\ \emph {et~al.}(1999)\citenamefont
  {Bouwmeester}, \citenamefont {Pan}, \citenamefont {Daniell}, \citenamefont
  {Weinfurter},\ and\ \citenamefont {Zeilinger}}]{bouwmeester1999}%
  \BibitemOpen
  \bibfield  {author} {\bibinfo {author} {\bibfnamefont {D.}~\bibnamefont
  {Bouwmeester}}, \bibinfo {author} {\bibfnamefont {J.-W.}\ \bibnamefont
  {Pan}}, \bibinfo {author} {\bibfnamefont {M.}~\bibnamefont {Daniell}},
  \bibinfo {author} {\bibfnamefont {H.}~\bibnamefont {Weinfurter}},\ and\
  \bibinfo {author} {\bibfnamefont {A.}~\bibnamefont {Zeilinger}},\ }\bibfield
  {title} {\bibinfo {title} {Observation of {{Three}}-{{Photon
  Greenberger}}-{{Horne}}-{{Zeilinger Entanglement}}},\ }\href
  {https://doi.org/10.1103/PhysRevLett.82.1345} {\bibfield  {journal} {\bibinfo
   {journal} {Physical Review Letters}\ }\textbf {\bibinfo {volume} {82}},\
  \bibinfo {pages} {1345} (\bibinfo {year} {1999})}\BibitemShut {NoStop}%
\bibitem [{\citenamefont {Knill}\ \emph {et~al.}(2001)\citenamefont {Knill},
  \citenamefont {Laflamme},\ and\ \citenamefont {Milburn}}]{knill2001}%
  \BibitemOpen
  \bibfield  {author} {\bibinfo {author} {\bibfnamefont {E.}~\bibnamefont
  {Knill}}, \bibinfo {author} {\bibfnamefont {R.}~\bibnamefont {Laflamme}},\
  and\ \bibinfo {author} {\bibfnamefont {G.~J.}\ \bibnamefont {Milburn}},\
  }\bibfield  {title} {\bibinfo {title} {A scheme for efficient quantum
  computation with linear optics},\ }\href@noop {} {\bibfield  {journal}
  {\bibinfo  {journal} {nature}\ }\textbf {\bibinfo {volume} {409}},\ \bibinfo
  {pages} {46} (\bibinfo {year} {2001})}\BibitemShut {NoStop}%
\bibitem [{\citenamefont {Spring}\ \emph {et~al.}(2013)\citenamefont {Spring},
  \citenamefont {Metcalf}, \citenamefont {Humphreys}, \citenamefont
  {Kolthammer}, \citenamefont {Jin}, \citenamefont {Barbieri}, \citenamefont
  {Datta}, \citenamefont {{Thomas-Peter}}, \citenamefont {Langford},
  \citenamefont {Kundys}, \citenamefont {Gates}, \citenamefont {Smith},
  \citenamefont {Smith},\ and\ \citenamefont {Walmsley}}]{spring2013}%
  \BibitemOpen
  \bibfield  {author} {\bibinfo {author} {\bibfnamefont {J.~B.}\ \bibnamefont
  {Spring}}, \bibinfo {author} {\bibfnamefont {B.~J.}\ \bibnamefont {Metcalf}},
  \bibinfo {author} {\bibfnamefont {P.~C.}\ \bibnamefont {Humphreys}}, \bibinfo
  {author} {\bibfnamefont {W.~S.}\ \bibnamefont {Kolthammer}}, \bibinfo
  {author} {\bibfnamefont {X.-M.}\ \bibnamefont {Jin}}, \bibinfo {author}
  {\bibfnamefont {M.}~\bibnamefont {Barbieri}}, \bibinfo {author}
  {\bibfnamefont {A.}~\bibnamefont {Datta}}, \bibinfo {author} {\bibfnamefont
  {N.}~\bibnamefont {{Thomas-Peter}}}, \bibinfo {author} {\bibfnamefont
  {N.~K.}\ \bibnamefont {Langford}}, \bibinfo {author} {\bibfnamefont
  {D.}~\bibnamefont {Kundys}}, \bibinfo {author} {\bibfnamefont {J.~C.}\
  \bibnamefont {Gates}}, \bibinfo {author} {\bibfnamefont {B.~J.}\ \bibnamefont
  {Smith}}, \bibinfo {author} {\bibfnamefont {P.~G.~R.}\ \bibnamefont
  {Smith}},\ and\ \bibinfo {author} {\bibfnamefont {I.~A.}\ \bibnamefont
  {Walmsley}},\ }\bibfield  {title} {\bibinfo {title} {Boson {{Sampling}} on a
  {{Photonic Chip}}},\ }\href {https://doi.org/10.1126/science.1231692}
  {\bibfield  {journal} {\bibinfo  {journal} {Science}\ }\textbf {\bibinfo
  {volume} {339}},\ \bibinfo {pages} {798} (\bibinfo {year}
  {2013})}\BibitemShut {NoStop}%
\bibitem [{\citenamefont {Ewert}\ and\ \citenamefont {{van
  Loock}}(2014)}]{ewert2014}%
  \BibitemOpen
  \bibfield  {author} {\bibinfo {author} {\bibfnamefont {F.}~\bibnamefont
  {Ewert}}\ and\ \bibinfo {author} {\bibfnamefont {P.}~\bibnamefont {{van
  Loock}}},\ }\bibfield  {title} {\bibinfo {title} {\$3/4\$-{{Efficient Bell
  Measurement}} with {{Passive Linear Optics}} and {{Unentangled Ancillae}}},\
  }\href {https://doi.org/10.1103/PhysRevLett.113.140403} {\bibfield  {journal}
  {\bibinfo  {journal} {Physical Review Letters}\ }\textbf {\bibinfo {volume}
  {113}},\ \bibinfo {pages} {140403} (\bibinfo {year} {2014})}\BibitemShut
  {NoStop}%
\bibitem [{\citenamefont {Hong}\ \emph {et~al.}(1987)\citenamefont {Hong},
  \citenamefont {Ou},\ and\ \citenamefont {Mandel}}]{hong1987a}%
  \BibitemOpen
  \bibfield  {author} {\bibinfo {author} {\bibfnamefont {C.~K.}\ \bibnamefont
  {Hong}}, \bibinfo {author} {\bibfnamefont {Z.~Y.}\ \bibnamefont {Ou}},\ and\
  \bibinfo {author} {\bibfnamefont {L.}~\bibnamefont {Mandel}},\ }\bibfield
  {title} {\bibinfo {title} {Measurement of subpicosecond time intervals
  between two photons by interference},\ }\href
  {https://doi.org/10.1103/PhysRevLett.59.2044} {\bibfield  {journal} {\bibinfo
   {journal} {Physical Review Letters}\ }\textbf {\bibinfo {volume} {59}},\
  \bibinfo {pages} {2044} (\bibinfo {year} {1987})}\BibitemShut {NoStop}%
\bibitem [{\citenamefont {Pittman}\ \emph {et~al.}(1996)\citenamefont
  {Pittman}, \citenamefont {Strekalov}, \citenamefont {Migdall}, \citenamefont
  {Rubin}, \citenamefont {Sergienko},\ and\ \citenamefont
  {Shih}}]{pittman1996}%
  \BibitemOpen
  \bibfield  {author} {\bibinfo {author} {\bibfnamefont {T.~B.}\ \bibnamefont
  {Pittman}}, \bibinfo {author} {\bibfnamefont {D.~V.}\ \bibnamefont
  {Strekalov}}, \bibinfo {author} {\bibfnamefont {A.}~\bibnamefont {Migdall}},
  \bibinfo {author} {\bibfnamefont {M.~H.}\ \bibnamefont {Rubin}}, \bibinfo
  {author} {\bibfnamefont {A.~V.}\ \bibnamefont {Sergienko}},\ and\ \bibinfo
  {author} {\bibfnamefont {Y.~H.}\ \bibnamefont {Shih}},\ }\bibfield  {title}
  {\bibinfo {title} {Can {{Two}}-{{Photon Interference}} be {{Considered}} the
  {{Interference}} of {{Two Photons}}?},\ }\href
  {https://doi.org/10.1103/PhysRevLett.77.1917} {\bibfield  {journal} {\bibinfo
   {journal} {Physical Review Letters}\ }\textbf {\bibinfo {volume} {77}},\
  \bibinfo {pages} {1917} (\bibinfo {year} {1996})}\BibitemShut {NoStop}%
\bibitem [{\citenamefont {Raymer}\ \emph {et~al.}(2010)\citenamefont {Raymer},
  \citenamefont {{van Enk}}, \citenamefont {McKinstrie},\ and\ \citenamefont
  {McGuinness}}]{raymer2010}%
  \BibitemOpen
  \bibfield  {author} {\bibinfo {author} {\bibfnamefont {M.~G.}\ \bibnamefont
  {Raymer}}, \bibinfo {author} {\bibfnamefont {S.~J.}\ \bibnamefont {{van
  Enk}}}, \bibinfo {author} {\bibfnamefont {C.~J.}\ \bibnamefont
  {McKinstrie}},\ and\ \bibinfo {author} {\bibfnamefont {H.~J.}\ \bibnamefont
  {McGuinness}},\ }\bibfield  {title} {\bibinfo {title} {Interference of two
  photons of different color},\ }\href
  {https://doi.org/10.1016/j.optcom.2009.10.057} {\bibfield  {journal}
  {\bibinfo  {journal} {Optics Communications}\ }\textbf {\bibinfo {volume}
  {283}},\ \bibinfo {pages} {747} (\bibinfo {year} {2010})}\BibitemShut
  {NoStop}%
\bibitem [{\citenamefont {Zwerger}\ \emph {et~al.}(2012)\citenamefont
  {Zwerger}, \citenamefont {D{\"u}r},\ and\ \citenamefont
  {Briegel}}]{zwerger2012a}%
  \BibitemOpen
  \bibfield  {author} {\bibinfo {author} {\bibfnamefont {M.}~\bibnamefont
  {Zwerger}}, \bibinfo {author} {\bibfnamefont {W.}~\bibnamefont {D{\"u}r}},\
  and\ \bibinfo {author} {\bibfnamefont {H.~J.}\ \bibnamefont {Briegel}},\
  }\bibfield  {title} {\bibinfo {title} {Measurement-based quantum repeaters},\
  }\href {https://doi.org/10.1103/PhysRevA.85.062326} {\bibfield  {journal}
  {\bibinfo  {journal} {Physical Review A}\ }\textbf {\bibinfo {volume} {85}},\
  \bibinfo {pages} {062326} (\bibinfo {year} {2012})}\BibitemShut {NoStop}%
\bibitem [{\citenamefont {Azuma}\ \emph {et~al.}(2015)\citenamefont {Azuma},
  \citenamefont {Tamaki},\ and\ \citenamefont {Lo}}]{azuma2015}%
  \BibitemOpen
  \bibfield  {author} {\bibinfo {author} {\bibfnamefont {K.}~\bibnamefont
  {Azuma}}, \bibinfo {author} {\bibfnamefont {K.}~\bibnamefont {Tamaki}},\ and\
  \bibinfo {author} {\bibfnamefont {H.-K.}\ \bibnamefont {Lo}},\ }\bibfield
  {title} {\bibinfo {title} {All-photonic quantum repeaters},\ }\href
  {https://doi.org/10.1038/ncomms7787} {\bibfield  {journal} {\bibinfo
  {journal} {Nature Communications}\ }\textbf {\bibinfo {volume} {6}},\
  \bibinfo {pages} {6787} (\bibinfo {year} {2015})}\BibitemShut {NoStop}%
\bibitem [{\citenamefont {Lukens}\ and\ \citenamefont
  {Lougovski}(2017)}]{lukens2017}%
  \BibitemOpen
  \bibfield  {author} {\bibinfo {author} {\bibfnamefont {J.~M.}\ \bibnamefont
  {Lukens}}\ and\ \bibinfo {author} {\bibfnamefont {P.}~\bibnamefont
  {Lougovski}},\ }\bibfield  {title} {\bibinfo {title} {Frequency-encoded
  photonic qubits for scalable quantum information processing},\ }\href
  {https://doi.org/10.1364/OPTICA.4.000008} {\bibfield  {journal} {\bibinfo
  {journal} {Optica}\ }\textbf {\bibinfo {volume} {4}},\ \bibinfo {pages} {8}
  (\bibinfo {year} {2017})}\BibitemShut {NoStop}%
\bibitem [{\citenamefont {Joshi}\ \emph {et~al.}(2018)\citenamefont {Joshi},
  \citenamefont {Farsi}, \citenamefont {Clemmen}, \citenamefont {Ramelow},\
  and\ \citenamefont {Gaeta}}]{joshi2018a}%
  \BibitemOpen
  \bibfield  {author} {\bibinfo {author} {\bibfnamefont {C.}~\bibnamefont
  {Joshi}}, \bibinfo {author} {\bibfnamefont {A.}~\bibnamefont {Farsi}},
  \bibinfo {author} {\bibfnamefont {S.}~\bibnamefont {Clemmen}}, \bibinfo
  {author} {\bibfnamefont {S.}~\bibnamefont {Ramelow}},\ and\ \bibinfo {author}
  {\bibfnamefont {A.~L.}\ \bibnamefont {Gaeta}},\ }\bibfield  {title} {\bibinfo
  {title} {Frequency multiplexing for quasi-deterministic heralded
  single-photon sources},\ }\href {https://doi.org/10.1038/s41467-018-03254-4}
  {\bibfield  {journal} {\bibinfo  {journal} {Nature Communications}\ }\textbf
  {\bibinfo {volume} {9}},\ \bibinfo {pages} {1} (\bibinfo {year}
  {2018})}\BibitemShut {NoStop}%
\bibitem [{\citenamefont {Hiemstra}\ \emph {et~al.}(2019)\citenamefont
  {Hiemstra}, \citenamefont {Parker}, \citenamefont {Humphreys}, \citenamefont
  {Tiedau}, \citenamefont {Beck}, \citenamefont {Karpi{\'n}ski}, \citenamefont
  {Smith}, \citenamefont {Eckstein}, \citenamefont {Kolthammer},\ and\
  \citenamefont {Walmsley}}]{hiemstra2019}%
  \BibitemOpen
  \bibfield  {author} {\bibinfo {author} {\bibfnamefont {T.}~\bibnamefont
  {Hiemstra}}, \bibinfo {author} {\bibfnamefont {T.~F.}\ \bibnamefont
  {Parker}}, \bibinfo {author} {\bibfnamefont {P.~C.}\ \bibnamefont
  {Humphreys}}, \bibinfo {author} {\bibfnamefont {J.}~\bibnamefont {Tiedau}},
  \bibinfo {author} {\bibfnamefont {M.}~\bibnamefont {Beck}}, \bibinfo {author}
  {\bibfnamefont {M.}~\bibnamefont {Karpi{\'n}ski}}, \bibinfo {author}
  {\bibfnamefont {B.~J.}\ \bibnamefont {Smith}}, \bibinfo {author}
  {\bibfnamefont {A.}~\bibnamefont {Eckstein}}, \bibinfo {author}
  {\bibfnamefont {W.~S.}\ \bibnamefont {Kolthammer}},\ and\ \bibinfo {author}
  {\bibfnamefont {I.~A.}\ \bibnamefont {Walmsley}},\ }\bibfield  {title}
  {\bibinfo {title} {Pure {{Single Photons}} from {{Scalable Frequency
  Multiplexing}}},\ }\href@noop {} {\bibfield  {journal} {\bibinfo  {journal}
  {arXiv:1907.10355 [quant-ph]}\ } (\bibinfo {year} {2019})},\ \Eprint
  {https://arxiv.org/abs/1907.10355} {arXiv:1907.10355 [quant-ph]} \BibitemShut
  {NoStop}%
\bibitem [{\citenamefont {Reimer}\ \emph {et~al.}(2014)\citenamefont {Reimer},
  \citenamefont {Caspani}, \citenamefont {Clerici}, \citenamefont {Ferrera},
  \citenamefont {Kues}, \citenamefont {Peccianti}, \citenamefont {Pasquazi},
  \citenamefont {Razzari}, \citenamefont {Little}, \citenamefont {Chu},
  \citenamefont {Moss},\ and\ \citenamefont {Morandotti}}]{reimer2014}%
  \BibitemOpen
  \bibfield  {author} {\bibinfo {author} {\bibfnamefont {C.}~\bibnamefont
  {Reimer}}, \bibinfo {author} {\bibfnamefont {L.}~\bibnamefont {Caspani}},
  \bibinfo {author} {\bibfnamefont {M.}~\bibnamefont {Clerici}}, \bibinfo
  {author} {\bibfnamefont {M.}~\bibnamefont {Ferrera}}, \bibinfo {author}
  {\bibfnamefont {M.}~\bibnamefont {Kues}}, \bibinfo {author} {\bibfnamefont
  {M.}~\bibnamefont {Peccianti}}, \bibinfo {author} {\bibfnamefont
  {A.}~\bibnamefont {Pasquazi}}, \bibinfo {author} {\bibfnamefont
  {L.}~\bibnamefont {Razzari}}, \bibinfo {author} {\bibfnamefont {B.~E.}\
  \bibnamefont {Little}}, \bibinfo {author} {\bibfnamefont {S.~T.}\
  \bibnamefont {Chu}}, \bibinfo {author} {\bibfnamefont {D.~J.}\ \bibnamefont
  {Moss}},\ and\ \bibinfo {author} {\bibfnamefont {R.}~\bibnamefont
  {Morandotti}},\ }\bibfield  {title} {\bibinfo {title} {Integrated frequency
  comb source of heralded single photons},\ }\href
  {https://doi.org/10.1364/OE.22.006535} {\bibfield  {journal} {\bibinfo
  {journal} {Optics Express}\ }\textbf {\bibinfo {volume} {22}},\ \bibinfo
  {pages} {6535} (\bibinfo {year} {2014})}\BibitemShut {NoStop}%
\bibitem [{\citenamefont {Harris}\ \emph {et~al.}(2014)\citenamefont {Harris},
  \citenamefont {Grassani}, \citenamefont {Simbula}, \citenamefont {Pant},
  \citenamefont {Galli}, \citenamefont {{Baehr-Jones}}, \citenamefont
  {Hochberg}, \citenamefont {Englund}, \citenamefont {Bajoni},\ and\
  \citenamefont {Galland}}]{harris2014}%
  \BibitemOpen
  \bibfield  {author} {\bibinfo {author} {\bibfnamefont {N.~C.}\ \bibnamefont
  {Harris}}, \bibinfo {author} {\bibfnamefont {D.}~\bibnamefont {Grassani}},
  \bibinfo {author} {\bibfnamefont {A.}~\bibnamefont {Simbula}}, \bibinfo
  {author} {\bibfnamefont {M.}~\bibnamefont {Pant}}, \bibinfo {author}
  {\bibfnamefont {M.}~\bibnamefont {Galli}}, \bibinfo {author} {\bibfnamefont
  {T.}~\bibnamefont {{Baehr-Jones}}}, \bibinfo {author} {\bibfnamefont
  {M.}~\bibnamefont {Hochberg}}, \bibinfo {author} {\bibfnamefont
  {D.}~\bibnamefont {Englund}}, \bibinfo {author} {\bibfnamefont
  {D.}~\bibnamefont {Bajoni}},\ and\ \bibinfo {author} {\bibfnamefont
  {C.}~\bibnamefont {Galland}},\ }\bibfield  {title} {\bibinfo {title}
  {Integrated {{Source}} of {{Spectrally Filtered Correlated Photons}} for
  {{Large}}-{{Scale Quantum Photonic Systems}}},\ }\href
  {https://doi.org/10.1103/PhysRevX.4.041047} {\bibfield  {journal} {\bibinfo
  {journal} {Physical Review X}\ }\textbf {\bibinfo {volume} {4}},\ \bibinfo
  {pages} {041047} (\bibinfo {year} {2014})}\BibitemShut {NoStop}%
\bibitem [{\citenamefont {Ramelow}\ \emph {et~al.}(2015)\citenamefont
  {Ramelow}, \citenamefont {Farsi}, \citenamefont {Clemmen}, \citenamefont
  {Orquiza}, \citenamefont {Luke}, \citenamefont {Lipson},\ and\ \citenamefont
  {Gaeta}}]{ramelow2015}%
  \BibitemOpen
  \bibfield  {author} {\bibinfo {author} {\bibfnamefont {S.}~\bibnamefont
  {Ramelow}}, \bibinfo {author} {\bibfnamefont {A.}~\bibnamefont {Farsi}},
  \bibinfo {author} {\bibfnamefont {S.}~\bibnamefont {Clemmen}}, \bibinfo
  {author} {\bibfnamefont {D.}~\bibnamefont {Orquiza}}, \bibinfo {author}
  {\bibfnamefont {K.}~\bibnamefont {Luke}}, \bibinfo {author} {\bibfnamefont
  {M.}~\bibnamefont {Lipson}},\ and\ \bibinfo {author} {\bibfnamefont {A.~L.}\
  \bibnamefont {Gaeta}},\ }\bibfield  {title} {\bibinfo {title}
  {Silicon-{{Nitride Platform}} for {{Narrowband Entangled Photon
  Generation}}},\ }\href@noop {} {\bibfield  {journal} {\bibinfo  {journal}
  {arXiv:1508.04358 [physics, physics:quant-ph]}\ } (\bibinfo {year} {2015})},\
  \Eprint {https://arxiv.org/abs/1508.04358} {arXiv:1508.04358 [physics,
  physics:quant-ph]} \BibitemShut {NoStop}%
\bibitem [{\citenamefont {Jiang}\ \emph {et~al.}(2015)\citenamefont {Jiang},
  \citenamefont {Lu}, \citenamefont {Zhang}, \citenamefont {Painter},\ and\
  \citenamefont {Lin}}]{jiang2015}%
  \BibitemOpen
  \bibfield  {author} {\bibinfo {author} {\bibfnamefont {W.~C.}\ \bibnamefont
  {Jiang}}, \bibinfo {author} {\bibfnamefont {X.}~\bibnamefont {Lu}}, \bibinfo
  {author} {\bibfnamefont {J.}~\bibnamefont {Zhang}}, \bibinfo {author}
  {\bibfnamefont {O.}~\bibnamefont {Painter}},\ and\ \bibinfo {author}
  {\bibfnamefont {Q.}~\bibnamefont {Lin}},\ }\bibfield  {title} {\bibinfo
  {title} {Silicon-chip source of bright photon pairs},\ }\href
  {https://doi.org/10.1364/OE.23.020884} {\bibfield  {journal} {\bibinfo
  {journal} {Optics Express}\ }\textbf {\bibinfo {volume} {23}},\ \bibinfo
  {pages} {20884} (\bibinfo {year} {2015})}\BibitemShut {NoStop}%
\bibitem [{\citenamefont {Grassani}\ \emph {et~al.}(2015)\citenamefont
  {Grassani}, \citenamefont {Azzini}, \citenamefont {Liscidini}, \citenamefont
  {Galli}, \citenamefont {Strain}, \citenamefont {Sorel}, \citenamefont
  {Sipe},\ and\ \citenamefont {Bajoni}}]{grassani2015}%
  \BibitemOpen
  \bibfield  {author} {\bibinfo {author} {\bibfnamefont {D.}~\bibnamefont
  {Grassani}}, \bibinfo {author} {\bibfnamefont {S.}~\bibnamefont {Azzini}},
  \bibinfo {author} {\bibfnamefont {M.}~\bibnamefont {Liscidini}}, \bibinfo
  {author} {\bibfnamefont {M.}~\bibnamefont {Galli}}, \bibinfo {author}
  {\bibfnamefont {M.~J.}\ \bibnamefont {Strain}}, \bibinfo {author}
  {\bibfnamefont {M.}~\bibnamefont {Sorel}}, \bibinfo {author} {\bibfnamefont
  {J.~E.}\ \bibnamefont {Sipe}},\ and\ \bibinfo {author} {\bibfnamefont
  {D.}~\bibnamefont {Bajoni}},\ }\bibfield  {title} {\bibinfo {title}
  {Micrometer-scale integrated silicon source of time-energy entangled
  photons},\ }\href {https://doi.org/10.1364/OPTICA.2.000088} {\bibfield
  {journal} {\bibinfo  {journal} {Optica}\ }\textbf {\bibinfo {volume} {2}},\
  \bibinfo {pages} {88} (\bibinfo {year} {2015})}\BibitemShut {NoStop}%
\bibitem [{\citenamefont {Montaut}\ \emph {et~al.}(2017)\citenamefont
  {Montaut}, \citenamefont {Sansoni}, \citenamefont {{Meyer-Scott}},
  \citenamefont {Ricken}, \citenamefont {Quiring}, \citenamefont {Herrmann},\
  and\ \citenamefont {Silberhorn}}]{montaut2017}%
  \BibitemOpen
  \bibfield  {author} {\bibinfo {author} {\bibfnamefont {N.}~\bibnamefont
  {Montaut}}, \bibinfo {author} {\bibfnamefont {L.}~\bibnamefont {Sansoni}},
  \bibinfo {author} {\bibfnamefont {E.}~\bibnamefont {{Meyer-Scott}}}, \bibinfo
  {author} {\bibfnamefont {R.}~\bibnamefont {Ricken}}, \bibinfo {author}
  {\bibfnamefont {V.}~\bibnamefont {Quiring}}, \bibinfo {author} {\bibfnamefont
  {H.}~\bibnamefont {Herrmann}},\ and\ \bibinfo {author} {\bibfnamefont
  {C.}~\bibnamefont {Silberhorn}},\ }\bibfield  {title} {\bibinfo {title}
  {High-{{Efficiency Plug}}-and-{{Play Source}} of {{Heralded Single
  Photons}}},\ }\href {https://doi.org/10.1103/PhysRevApplied.8.024021}
  {\bibfield  {journal} {\bibinfo  {journal} {Physical Review Applied}\
  }\textbf {\bibinfo {volume} {8}},\ \bibinfo {pages} {024021} (\bibinfo {year}
  {2017})}\BibitemShut {NoStop}%
\bibitem [{\citenamefont {{Meyer-Scott}}\ \emph {et~al.}(2018)\citenamefont
  {{Meyer-Scott}}, \citenamefont {Prasannan}, \citenamefont {Eigner},
  \citenamefont {Quiring}, \citenamefont {Donohue}, \citenamefont {Barkhofen},\
  and\ \citenamefont {Silberhorn}}]{meyer-scott2018}%
  \BibitemOpen
  \bibfield  {author} {\bibinfo {author} {\bibfnamefont {E.}~\bibnamefont
  {{Meyer-Scott}}}, \bibinfo {author} {\bibfnamefont {N.}~\bibnamefont
  {Prasannan}}, \bibinfo {author} {\bibfnamefont {C.}~\bibnamefont {Eigner}},
  \bibinfo {author} {\bibfnamefont {V.}~\bibnamefont {Quiring}}, \bibinfo
  {author} {\bibfnamefont {J.~M.}\ \bibnamefont {Donohue}}, \bibinfo {author}
  {\bibfnamefont {S.}~\bibnamefont {Barkhofen}},\ and\ \bibinfo {author}
  {\bibfnamefont {C.}~\bibnamefont {Silberhorn}},\ }\bibfield  {title}
  {\bibinfo {title} {High-performance source of spectrally pure, polarization
  entangled photon pairs based on hybrid integrated-bulk optics},\ }\href
  {https://doi.org/10.1364/OE.26.032475} {\bibfield  {journal} {\bibinfo
  {journal} {Optics Express}\ }\textbf {\bibinfo {volume} {26}},\ \bibinfo
  {pages} {32475} (\bibinfo {year} {2018})}\BibitemShut {NoStop}%
\bibitem [{\citenamefont {Qiang}\ \emph {et~al.}(2018)\citenamefont {Qiang},
  \citenamefont {Zhou}, \citenamefont {Wang}, \citenamefont {Wilkes},
  \citenamefont {Loke}, \citenamefont {O'Gara}, \citenamefont {Kling},
  \citenamefont {Marshall}, \citenamefont {Santagati}, \citenamefont {Ralph},
  \citenamefont {Wang}, \citenamefont {O'Brien}, \citenamefont {Thompson},\
  and\ \citenamefont {Matthews}}]{qiang2018}%
  \BibitemOpen
  \bibfield  {author} {\bibinfo {author} {\bibfnamefont {X.}~\bibnamefont
  {Qiang}}, \bibinfo {author} {\bibfnamefont {X.}~\bibnamefont {Zhou}},
  \bibinfo {author} {\bibfnamefont {J.}~\bibnamefont {Wang}}, \bibinfo {author}
  {\bibfnamefont {C.~M.}\ \bibnamefont {Wilkes}}, \bibinfo {author}
  {\bibfnamefont {T.}~\bibnamefont {Loke}}, \bibinfo {author} {\bibfnamefont
  {S.}~\bibnamefont {O'Gara}}, \bibinfo {author} {\bibfnamefont
  {L.}~\bibnamefont {Kling}}, \bibinfo {author} {\bibfnamefont {G.~D.}\
  \bibnamefont {Marshall}}, \bibinfo {author} {\bibfnamefont {R.}~\bibnamefont
  {Santagati}}, \bibinfo {author} {\bibfnamefont {T.~C.}\ \bibnamefont
  {Ralph}}, \bibinfo {author} {\bibfnamefont {J.~B.}\ \bibnamefont {Wang}},
  \bibinfo {author} {\bibfnamefont {J.~L.}\ \bibnamefont {O'Brien}}, \bibinfo
  {author} {\bibfnamefont {M.~G.}\ \bibnamefont {Thompson}},\ and\ \bibinfo
  {author} {\bibfnamefont {J.~C.~F.}\ \bibnamefont {Matthews}},\ }\bibfield
  {title} {\bibinfo {title} {Large-scale silicon quantum photonics implementing
  arbitrary two-qubit processing},\ }\href
  {https://doi.org/10.1038/s41566-018-0236-y} {\bibfield  {journal} {\bibinfo
  {journal} {Nature Photonics}\ }\textbf {\bibinfo {volume} {12}},\ \bibinfo
  {pages} {534} (\bibinfo {year} {2018})}\BibitemShut {NoStop}%
\bibitem [{\citenamefont {Paesani}\ \emph {et~al.}(2019)\citenamefont
  {Paesani}, \citenamefont {Ding}, \citenamefont {Santagati}, \citenamefont
  {Chakhmakhchyan}, \citenamefont {Vigliar}, \citenamefont {Rottwitt},
  \citenamefont {Oxenl{\o}we}, \citenamefont {Wang}, \citenamefont {Thompson},\
  and\ \citenamefont {Laing}}]{paesani2019}%
  \BibitemOpen
  \bibfield  {author} {\bibinfo {author} {\bibfnamefont {S.}~\bibnamefont
  {Paesani}}, \bibinfo {author} {\bibfnamefont {Y.}~\bibnamefont {Ding}},
  \bibinfo {author} {\bibfnamefont {R.}~\bibnamefont {Santagati}}, \bibinfo
  {author} {\bibfnamefont {L.}~\bibnamefont {Chakhmakhchyan}}, \bibinfo
  {author} {\bibfnamefont {C.}~\bibnamefont {Vigliar}}, \bibinfo {author}
  {\bibfnamefont {K.}~\bibnamefont {Rottwitt}}, \bibinfo {author}
  {\bibfnamefont {L.~K.}\ \bibnamefont {Oxenl{\o}we}}, \bibinfo {author}
  {\bibfnamefont {J.}~\bibnamefont {Wang}}, \bibinfo {author} {\bibfnamefont
  {M.~G.}\ \bibnamefont {Thompson}},\ and\ \bibinfo {author} {\bibfnamefont
  {A.}~\bibnamefont {Laing}},\ }\bibfield  {title} {\bibinfo {title}
  {Generation and sampling of quantum states of light in a silicon chip},\
  }\href {https://doi.org/10.1038/s41567-019-0567-8} {\bibfield  {journal}
  {\bibinfo  {journal} {Nature Physics}\ }\textbf {\bibinfo {volume} {15}},\
  \bibinfo {pages} {925} (\bibinfo {year} {2019})}\BibitemShut {NoStop}%
\bibitem [{\citenamefont {McGuinness}\ \emph {et~al.}(2010)\citenamefont
  {McGuinness}, \citenamefont {Raymer}, \citenamefont {McKinstrie},\ and\
  \citenamefont {Radic}}]{mcguinness2010}%
  \BibitemOpen
  \bibfield  {author} {\bibinfo {author} {\bibfnamefont {H.~J.}\ \bibnamefont
  {McGuinness}}, \bibinfo {author} {\bibfnamefont {M.~G.}\ \bibnamefont
  {Raymer}}, \bibinfo {author} {\bibfnamefont {C.~J.}\ \bibnamefont
  {McKinstrie}},\ and\ \bibinfo {author} {\bibfnamefont {S.}~\bibnamefont
  {Radic}},\ }\bibfield  {title} {\bibinfo {title} {Quantum {{Frequency
  Translation}} of {{Single}}-{{Photon States}} in a {{Photonic Crystal
  Fiber}}},\ }\href {https://doi.org/10.1103/PhysRevLett.105.093604} {\bibfield
   {journal} {\bibinfo  {journal} {Physical Review Letters}\ }\textbf {\bibinfo
  {volume} {105}},\ \bibinfo {pages} {093604} (\bibinfo {year}
  {2010})}\BibitemShut {NoStop}%
\bibitem [{\citenamefont {Farsi}(2015)}]{farsi2015a}%
  \BibitemOpen
  \bibfield  {author} {\bibinfo {author} {\bibfnamefont {A.}~\bibnamefont
  {Farsi}},\ }\emph {\bibinfo {title} {Coherent {{Manipulation Of Light In The
  Classical And Quantum Regimes Via Four}}-{{Wave Mixing Bragg Scattering}}}},\
  \href@noop {} {Ph.D. thesis},\ \bibinfo  {school} {Cornell University}
  (\bibinfo {year} {2015})\BibitemShut {NoStop}%
\bibitem [{\citenamefont {McKinstrie}\ \emph {et~al.}(2004)\citenamefont
  {McKinstrie}, \citenamefont {Radic},\ and\ \citenamefont
  {Raymer}}]{mckinstrie2004}%
  \BibitemOpen
  \bibfield  {author} {\bibinfo {author} {\bibfnamefont {C.~J.}\ \bibnamefont
  {McKinstrie}}, \bibinfo {author} {\bibfnamefont {S.}~\bibnamefont {Radic}},\
  and\ \bibinfo {author} {\bibfnamefont {M.~G.}\ \bibnamefont {Raymer}},\
  }\bibfield  {title} {\bibinfo {title} {Quantum noise properties of parametric
  amplifiers driven by two pump waves},\ }\href
  {https://doi.org/10.1364/OPEX.12.005037} {\bibfield  {journal} {\bibinfo
  {journal} {Optics Express}\ }\textbf {\bibinfo {volume} {12}},\ \bibinfo
  {pages} {5037} (\bibinfo {year} {2004})}\BibitemShut {NoStop}%
\bibitem [{\citenamefont {McKinstrie}\ \emph
  {et~al.}(2005{\natexlab{a}})\citenamefont {McKinstrie}, \citenamefont
  {Harvey}, \citenamefont {Radic},\ and\ \citenamefont
  {Raymer}}]{mckinstrie2005a}%
  \BibitemOpen
  \bibfield  {author} {\bibinfo {author} {\bibfnamefont {C.~J.}\ \bibnamefont
  {McKinstrie}}, \bibinfo {author} {\bibfnamefont {J.~D.}\ \bibnamefont
  {Harvey}}, \bibinfo {author} {\bibfnamefont {S.}~\bibnamefont {Radic}},\ and\
  \bibinfo {author} {\bibfnamefont {M.~G.}\ \bibnamefont {Raymer}},\ }\bibfield
   {title} {\bibinfo {title} {Translation of quantum states by four-wave mixing
  in fibers},\ }\href {https://doi.org/10.1364/OPEX.13.009131} {\bibfield
  {journal} {\bibinfo  {journal} {Optics Express}\ }\textbf {\bibinfo {volume}
  {13}},\ \bibinfo {pages} {9131} (\bibinfo {year}
  {2005}{\natexlab{a}})}\BibitemShut {NoStop}%
\bibitem [{\citenamefont {McKinstrie}\ \emph
  {et~al.}(2005{\natexlab{b}})\citenamefont {McKinstrie}, \citenamefont {Yu},
  \citenamefont {Raymer},\ and\ \citenamefont {Radic}}]{mckinstrie2005}%
  \BibitemOpen
  \bibfield  {author} {\bibinfo {author} {\bibfnamefont {C.~J.}\ \bibnamefont
  {McKinstrie}}, \bibinfo {author} {\bibfnamefont {M.}~\bibnamefont {Yu}},
  \bibinfo {author} {\bibfnamefont {M.~G.}\ \bibnamefont {Raymer}},\ and\
  \bibinfo {author} {\bibfnamefont {S.}~\bibnamefont {Radic}},\ }\bibfield
  {title} {\bibinfo {title} {Quantum noise properties of parametric
  processes},\ }\href {https://doi.org/10.1364/OPEX.13.004986} {\bibfield
  {journal} {\bibinfo  {journal} {Optics Express}\ }\textbf {\bibinfo {volume}
  {13}},\ \bibinfo {pages} {4986} (\bibinfo {year}
  {2005}{\natexlab{b}})}\BibitemShut {NoStop}%
\bibitem [{\citenamefont {Clemmen}\ \emph {et~al.}(2016)\citenamefont
  {Clemmen}, \citenamefont {Farsi}, \citenamefont {Ramelow},\ and\
  \citenamefont {Gaeta}}]{clemmen2016}%
  \BibitemOpen
  \bibfield  {author} {\bibinfo {author} {\bibfnamefont {S.}~\bibnamefont
  {Clemmen}}, \bibinfo {author} {\bibfnamefont {A.}~\bibnamefont {Farsi}},
  \bibinfo {author} {\bibfnamefont {S.}~\bibnamefont {Ramelow}},\ and\ \bibinfo
  {author} {\bibfnamefont {A.~L.}\ \bibnamefont {Gaeta}},\ }\bibfield  {title}
  {\bibinfo {title} {Ramsey {{Interference}} with {{Single Photons}}},\ }\href
  {https://doi.org/10.1103/PhysRevLett.117.223601} {\bibfield  {journal}
  {\bibinfo  {journal} {Physical Review Letters}\ }\textbf {\bibinfo {volume}
  {117}},\ \bibinfo {pages} {223601} (\bibinfo {year} {2016})}\BibitemShut
  {NoStop}%
\bibitem [{\citenamefont {Kues}\ \emph {et~al.}(2017)\citenamefont {Kues},
  \citenamefont {Reimer}, \citenamefont {Roztocki}, \citenamefont {Cort{\'e}s},
  \citenamefont {Sciara}, \citenamefont {Wetzel}, \citenamefont {Zhang},
  \citenamefont {Cino}, \citenamefont {Chu}, \citenamefont {Little},
  \citenamefont {Moss}, \citenamefont {Caspani}, \citenamefont {Aza{\~n}a},\
  and\ \citenamefont {Morandotti}}]{kues2017}%
  \BibitemOpen
  \bibfield  {author} {\bibinfo {author} {\bibfnamefont {M.}~\bibnamefont
  {Kues}}, \bibinfo {author} {\bibfnamefont {C.}~\bibnamefont {Reimer}},
  \bibinfo {author} {\bibfnamefont {P.}~\bibnamefont {Roztocki}}, \bibinfo
  {author} {\bibfnamefont {L.~R.}\ \bibnamefont {Cort{\'e}s}}, \bibinfo
  {author} {\bibfnamefont {S.}~\bibnamefont {Sciara}}, \bibinfo {author}
  {\bibfnamefont {B.}~\bibnamefont {Wetzel}}, \bibinfo {author} {\bibfnamefont
  {Y.}~\bibnamefont {Zhang}}, \bibinfo {author} {\bibfnamefont
  {A.}~\bibnamefont {Cino}}, \bibinfo {author} {\bibfnamefont {S.~T.}\
  \bibnamefont {Chu}}, \bibinfo {author} {\bibfnamefont {B.~E.}\ \bibnamefont
  {Little}}, \bibinfo {author} {\bibfnamefont {D.~J.}\ \bibnamefont {Moss}},
  \bibinfo {author} {\bibfnamefont {L.}~\bibnamefont {Caspani}}, \bibinfo
  {author} {\bibfnamefont {J.}~\bibnamefont {Aza{\~n}a}},\ and\ \bibinfo
  {author} {\bibfnamefont {R.}~\bibnamefont {Morandotti}},\ }\bibfield  {title}
  {\bibinfo {title} {On-chip generation of high-dimensional entangled quantum
  states and their coherent control},\ }\href
  {https://doi.org/10.1038/nature22986} {\bibfield  {journal} {\bibinfo
  {journal} {Nature}\ }\textbf {\bibinfo {volume} {546}},\ \bibinfo {pages}
  {622} (\bibinfo {year} {2017})}\BibitemShut {NoStop}%
\bibitem [{\citenamefont {Reimer}\ \emph {et~al.}(2019)\citenamefont {Reimer},
  \citenamefont {Sciara}, \citenamefont {Roztocki}, \citenamefont {Islam},
  \citenamefont {Cort{\'e}s}, \citenamefont {Zhang}, \citenamefont {Fischer},
  \citenamefont {Loranger}, \citenamefont {Kashyap}, \citenamefont {Cino},
  \citenamefont {Chu}, \citenamefont {Little}, \citenamefont {Moss},
  \citenamefont {Caspani}, \citenamefont {Munro}, \citenamefont {Aza{\~n}a},
  \citenamefont {Kues},\ and\ \citenamefont {Morandotti}}]{reimer2019}%
  \BibitemOpen
  \bibfield  {author} {\bibinfo {author} {\bibfnamefont {C.}~\bibnamefont
  {Reimer}}, \bibinfo {author} {\bibfnamefont {S.}~\bibnamefont {Sciara}},
  \bibinfo {author} {\bibfnamefont {P.}~\bibnamefont {Roztocki}}, \bibinfo
  {author} {\bibfnamefont {M.}~\bibnamefont {Islam}}, \bibinfo {author}
  {\bibfnamefont {L.~R.}\ \bibnamefont {Cort{\'e}s}}, \bibinfo {author}
  {\bibfnamefont {Y.}~\bibnamefont {Zhang}}, \bibinfo {author} {\bibfnamefont
  {B.}~\bibnamefont {Fischer}}, \bibinfo {author} {\bibfnamefont
  {S.}~\bibnamefont {Loranger}}, \bibinfo {author} {\bibfnamefont
  {R.}~\bibnamefont {Kashyap}}, \bibinfo {author} {\bibfnamefont
  {A.}~\bibnamefont {Cino}}, \bibinfo {author} {\bibfnamefont {S.~T.}\
  \bibnamefont {Chu}}, \bibinfo {author} {\bibfnamefont {B.~E.}\ \bibnamefont
  {Little}}, \bibinfo {author} {\bibfnamefont {D.~J.}\ \bibnamefont {Moss}},
  \bibinfo {author} {\bibfnamefont {L.}~\bibnamefont {Caspani}}, \bibinfo
  {author} {\bibfnamefont {W.~J.}\ \bibnamefont {Munro}}, \bibinfo {author}
  {\bibfnamefont {J.}~\bibnamefont {Aza{\~n}a}}, \bibinfo {author}
  {\bibfnamefont {M.}~\bibnamefont {Kues}},\ and\ \bibinfo {author}
  {\bibfnamefont {R.}~\bibnamefont {Morandotti}},\ }\bibfield  {title}
  {\bibinfo {title} {High-dimensional one-way quantum processing implemented on
  d -level cluster states},\ }\href {https://doi.org/10.1038/s41567-018-0347-x}
  {\bibfield  {journal} {\bibinfo  {journal} {Nature Physics}\ }\textbf
  {\bibinfo {volume} {15}},\ \bibinfo {pages} {148} (\bibinfo {year}
  {2019})}\BibitemShut {NoStop}%
\bibitem [{\citenamefont {Imany}\ \emph {et~al.}(2019)\citenamefont {Imany},
  \citenamefont {{Jaramillo-Villegas}}, \citenamefont {Alshaykh}, \citenamefont
  {Lukens}, \citenamefont {Odele}, \citenamefont {Moore}, \citenamefont
  {Leaird}, \citenamefont {Qi},\ and\ \citenamefont {Weiner}}]{imany2019}%
  \BibitemOpen
  \bibfield  {author} {\bibinfo {author} {\bibfnamefont {P.}~\bibnamefont
  {Imany}}, \bibinfo {author} {\bibfnamefont {J.~A.}\ \bibnamefont
  {{Jaramillo-Villegas}}}, \bibinfo {author} {\bibfnamefont {M.~S.}\
  \bibnamefont {Alshaykh}}, \bibinfo {author} {\bibfnamefont {J.~M.}\
  \bibnamefont {Lukens}}, \bibinfo {author} {\bibfnamefont {O.~D.}\
  \bibnamefont {Odele}}, \bibinfo {author} {\bibfnamefont {A.~J.}\ \bibnamefont
  {Moore}}, \bibinfo {author} {\bibfnamefont {D.~E.}\ \bibnamefont {Leaird}},
  \bibinfo {author} {\bibfnamefont {M.}~\bibnamefont {Qi}},\ and\ \bibinfo
  {author} {\bibfnamefont {A.~M.}\ \bibnamefont {Weiner}},\ }\bibfield  {title}
  {\bibinfo {title} {High-dimensional optical quantum logic in large
  operational spaces},\ }\href {https://doi.org/10.1038/s41534-019-0173-8}
  {\bibfield  {journal} {\bibinfo  {journal} {npj Quantum Information}\
  }\textbf {\bibinfo {volume} {5}},\ \bibinfo {pages} {1} (\bibinfo {year}
  {2019})}\BibitemShut {NoStop}%
\bibitem [{\citenamefont {M{\'e}rolla}\ \emph {et~al.}(1999)\citenamefont
  {M{\'e}rolla}, \citenamefont {Mazurenko}, \citenamefont {Goedgebuer},\ and\
  \citenamefont {Rhodes}}]{merolla1999a}%
  \BibitemOpen
  \bibfield  {author} {\bibinfo {author} {\bibfnamefont {J.-M.}\ \bibnamefont
  {M{\'e}rolla}}, \bibinfo {author} {\bibfnamefont {Y.}~\bibnamefont
  {Mazurenko}}, \bibinfo {author} {\bibfnamefont {J.-P.}\ \bibnamefont
  {Goedgebuer}},\ and\ \bibinfo {author} {\bibfnamefont {W.~T.}\ \bibnamefont
  {Rhodes}},\ }\bibfield  {title} {\bibinfo {title} {Single-{{Photon
  Interference}} in {{Sidebands}} of {{Phase}}-{{Modulated Light}} for
  {{Quantum Cryptography}}},\ }\href
  {https://doi.org/10.1103/PhysRevLett.82.1656} {\bibfield  {journal} {\bibinfo
   {journal} {Physical Review Letters}\ }\textbf {\bibinfo {volume} {82}},\
  \bibinfo {pages} {1656} (\bibinfo {year} {1999})}\BibitemShut {NoStop}%
\bibitem [{\citenamefont {Olislager}\ \emph {et~al.}(2010)\citenamefont
  {Olislager}, \citenamefont {Cussey}, \citenamefont {Nguyen}, \citenamefont
  {Emplit}, \citenamefont {Massar}, \citenamefont {Merolla},\ and\
  \citenamefont {Huy}}]{olislager2010}%
  \BibitemOpen
  \bibfield  {author} {\bibinfo {author} {\bibfnamefont {L.}~\bibnamefont
  {Olislager}}, \bibinfo {author} {\bibfnamefont {J.}~\bibnamefont {Cussey}},
  \bibinfo {author} {\bibfnamefont {A.~T.}\ \bibnamefont {Nguyen}}, \bibinfo
  {author} {\bibfnamefont {P.}~\bibnamefont {Emplit}}, \bibinfo {author}
  {\bibfnamefont {S.}~\bibnamefont {Massar}}, \bibinfo {author} {\bibfnamefont
  {J.-M.}\ \bibnamefont {Merolla}},\ and\ \bibinfo {author} {\bibfnamefont
  {K.~P.}\ \bibnamefont {Huy}},\ }\bibfield  {title} {\bibinfo {title}
  {Frequency-bin entangled photons},\ }\href
  {https://doi.org/10.1103/PhysRevA.82.013804} {\bibfield  {journal} {\bibinfo
  {journal} {Physical Review A}\ }\textbf {\bibinfo {volume} {82}},\ \bibinfo
  {pages} {013804} (\bibinfo {year} {2010})}\BibitemShut {NoStop}%
\bibitem [{\citenamefont {Olislager}\ \emph {et~al.}(2014)\citenamefont
  {Olislager}, \citenamefont {Woodhead}, \citenamefont {Phan~Huy},
  \citenamefont {Merolla}, \citenamefont {Emplit},\ and\ \citenamefont
  {Massar}}]{olislager2014}%
  \BibitemOpen
  \bibfield  {author} {\bibinfo {author} {\bibfnamefont {L.}~\bibnamefont
  {Olislager}}, \bibinfo {author} {\bibfnamefont {E.}~\bibnamefont {Woodhead}},
  \bibinfo {author} {\bibfnamefont {K.}~\bibnamefont {Phan~Huy}}, \bibinfo
  {author} {\bibfnamefont {J.-M.}\ \bibnamefont {Merolla}}, \bibinfo {author}
  {\bibfnamefont {P.}~\bibnamefont {Emplit}},\ and\ \bibinfo {author}
  {\bibfnamefont {S.}~\bibnamefont {Massar}},\ }\bibfield  {title} {\bibinfo
  {title} {Creating and manipulating entangled optical qubits in the frequency
  domain},\ }\href {https://doi.org/10.1103/PhysRevA.89.052323} {\bibfield
  {journal} {\bibinfo  {journal} {Physical Review A}\ }\textbf {\bibinfo
  {volume} {89}},\ \bibinfo {pages} {052323} (\bibinfo {year}
  {2014})}\BibitemShut {NoStop}%
\bibitem [{\citenamefont {Lu}\ \emph {et~al.}(2018{\natexlab{a}})\citenamefont
  {Lu}, \citenamefont {Lukens}, \citenamefont {Peters}, \citenamefont
  {Williams}, \citenamefont {Weiner},\ and\ \citenamefont
  {Lougovski}}]{lu2018}%
  \BibitemOpen
  \bibfield  {author} {\bibinfo {author} {\bibfnamefont {H.-H.}\ \bibnamefont
  {Lu}}, \bibinfo {author} {\bibfnamefont {J.~M.}\ \bibnamefont {Lukens}},
  \bibinfo {author} {\bibfnamefont {N.~A.}\ \bibnamefont {Peters}}, \bibinfo
  {author} {\bibfnamefont {B.~P.}\ \bibnamefont {Williams}}, \bibinfo {author}
  {\bibfnamefont {A.~M.}\ \bibnamefont {Weiner}},\ and\ \bibinfo {author}
  {\bibfnamefont {P.}~\bibnamefont {Lougovski}},\ }\bibfield  {title} {\bibinfo
  {title} {Quantum interference and correlation control of frequency-bin
  qubits},\ }\href {https://doi.org/10.1364/OPTICA.5.001455} {\bibfield
  {journal} {\bibinfo  {journal} {Optica}\ }\textbf {\bibinfo {volume} {5}},\
  \bibinfo {pages} {1455} (\bibinfo {year} {2018}{\natexlab{a}})}\BibitemShut
  {NoStop}%
\bibitem [{\citenamefont {Imany}\ \emph {et~al.}(2018)\citenamefont {Imany},
  \citenamefont {Odele}, \citenamefont {Alshaykh}, \citenamefont {Lu},
  \citenamefont {Leaird},\ and\ \citenamefont {Weiner}}]{imany2018}%
  \BibitemOpen
  \bibfield  {author} {\bibinfo {author} {\bibfnamefont {P.}~\bibnamefont
  {Imany}}, \bibinfo {author} {\bibfnamefont {O.~D.}\ \bibnamefont {Odele}},
  \bibinfo {author} {\bibfnamefont {M.~S.}\ \bibnamefont {Alshaykh}}, \bibinfo
  {author} {\bibfnamefont {H.-H.}\ \bibnamefont {Lu}}, \bibinfo {author}
  {\bibfnamefont {D.~E.}\ \bibnamefont {Leaird}},\ and\ \bibinfo {author}
  {\bibfnamefont {A.~M.}\ \bibnamefont {Weiner}},\ }\bibfield  {title}
  {\bibinfo {title} {Frequency-domain {{Hong}}-{{Ou}}-{{Mandel}} interference
  with linear optics},\ }\href {https://doi.org/10.1364/OL.43.002760}
  {\bibfield  {journal} {\bibinfo  {journal} {Optics Letters}\ }\textbf
  {\bibinfo {volume} {43}},\ \bibinfo {pages} {2760} (\bibinfo {year}
  {2018})}\BibitemShut {NoStop}%
\bibitem [{\citenamefont {Lu}\ \emph {et~al.}(2018{\natexlab{b}})\citenamefont
  {Lu}, \citenamefont {Lukens}, \citenamefont {Peters}, \citenamefont {Odele},
  \citenamefont {Leaird}, \citenamefont {Weiner},\ and\ \citenamefont
  {Lougovski}}]{lu2018a}%
  \BibitemOpen
  \bibfield  {author} {\bibinfo {author} {\bibfnamefont {H.-H.}\ \bibnamefont
  {Lu}}, \bibinfo {author} {\bibfnamefont {J.~M.}\ \bibnamefont {Lukens}},
  \bibinfo {author} {\bibfnamefont {N.~A.}\ \bibnamefont {Peters}}, \bibinfo
  {author} {\bibfnamefont {O.~D.}\ \bibnamefont {Odele}}, \bibinfo {author}
  {\bibfnamefont {D.~E.}\ \bibnamefont {Leaird}}, \bibinfo {author}
  {\bibfnamefont {A.~M.}\ \bibnamefont {Weiner}},\ and\ \bibinfo {author}
  {\bibfnamefont {P.}~\bibnamefont {Lougovski}},\ }\bibfield  {title} {\bibinfo
  {title} {Electro-{{Optic Frequency Beam Splitters}} and {{Tritters}} for
  {{High}}-{{Fidelity Photonic Quantum Information Processing}}},\ }\href
  {https://doi.org/10.1103/PhysRevLett.120.030502} {\bibfield  {journal}
  {\bibinfo  {journal} {Physical Review Letters}\ }\textbf {\bibinfo {volume}
  {120}},\ \bibinfo {pages} {030502} (\bibinfo {year}
  {2018}{\natexlab{b}})}\BibitemShut {NoStop}%
\bibitem [{\citenamefont {Galm{\`e}s}\ \emph {et~al.}(2019)\citenamefont
  {Galm{\`e}s}, \citenamefont {{Phan-Huy}}, \citenamefont {Furfaro},
  \citenamefont {Chembo},\ and\ \citenamefont {Merolla}}]{galmes2019a}%
  \BibitemOpen
  \bibfield  {author} {\bibinfo {author} {\bibfnamefont {B.}~\bibnamefont
  {Galm{\`e}s}}, \bibinfo {author} {\bibfnamefont {K.}~\bibnamefont
  {{Phan-Huy}}}, \bibinfo {author} {\bibfnamefont {L.}~\bibnamefont {Furfaro}},
  \bibinfo {author} {\bibfnamefont {Y.~K.}\ \bibnamefont {Chembo}},\ and\
  \bibinfo {author} {\bibfnamefont {J.-M.}\ \bibnamefont {Merolla}},\
  }\bibfield  {title} {\bibinfo {title} {Nine-frequency-path quantum
  interferometry over 60 km of optical fiber},\ }\href
  {https://doi.org/10.1103/PhysRevA.99.033805} {\bibfield  {journal} {\bibinfo
  {journal} {Physical Review A}\ }\textbf {\bibinfo {volume} {99}},\ \bibinfo
  {pages} {033805} (\bibinfo {year} {2019})}\BibitemShut {NoStop}%
\bibitem [{\citenamefont {Kobayashi}\ \emph {et~al.}(2016)\citenamefont
  {Kobayashi}, \citenamefont {Ikuta}, \citenamefont {Yasui}, \citenamefont
  {Miki}, \citenamefont {Yamashita}, \citenamefont {Terai}, \citenamefont
  {Yamamoto}, \citenamefont {Koashi},\ and\ \citenamefont
  {Imoto}}]{kobayashi2016}%
  \BibitemOpen
  \bibfield  {author} {\bibinfo {author} {\bibfnamefont {T.}~\bibnamefont
  {Kobayashi}}, \bibinfo {author} {\bibfnamefont {R.}~\bibnamefont {Ikuta}},
  \bibinfo {author} {\bibfnamefont {S.}~\bibnamefont {Yasui}}, \bibinfo
  {author} {\bibfnamefont {S.}~\bibnamefont {Miki}}, \bibinfo {author}
  {\bibfnamefont {T.}~\bibnamefont {Yamashita}}, \bibinfo {author}
  {\bibfnamefont {H.}~\bibnamefont {Terai}}, \bibinfo {author} {\bibfnamefont
  {T.}~\bibnamefont {Yamamoto}}, \bibinfo {author} {\bibfnamefont
  {M.}~\bibnamefont {Koashi}},\ and\ \bibinfo {author} {\bibfnamefont
  {N.}~\bibnamefont {Imoto}},\ }\bibfield  {title} {\bibinfo {title}
  {Frequency-domain {{Hong}}\textendash{{Ou}}\textendash{{Mandel}}
  interference},\ }\href {https://doi.org/10.1038/nphoton.2016.74} {\bibfield
  {journal} {\bibinfo  {journal} {Nature Photonics}\ }\textbf {\bibinfo
  {volume} {10}},\ \bibinfo {pages} {441} (\bibinfo {year} {2016})}\BibitemShut
  {NoStop}%
\bibitem [{\citenamefont {Vernon}\ \emph {et~al.}(2017)\citenamefont {Vernon},
  \citenamefont {Menotti}, \citenamefont {Tison}, \citenamefont {Steidle},
  \citenamefont {Fanto}, \citenamefont {Thomas}, \citenamefont {Preble},
  \citenamefont {Smith}, \citenamefont {Alsing}, \citenamefont {Liscidini},\
  and\ \citenamefont {Sipe}}]{vernon2017}%
  \BibitemOpen
  \bibfield  {author} {\bibinfo {author} {\bibfnamefont {Z.}~\bibnamefont
  {Vernon}}, \bibinfo {author} {\bibfnamefont {M.}~\bibnamefont {Menotti}},
  \bibinfo {author} {\bibfnamefont {C.~C.}\ \bibnamefont {Tison}}, \bibinfo
  {author} {\bibfnamefont {J.~A.}\ \bibnamefont {Steidle}}, \bibinfo {author}
  {\bibfnamefont {M.~L.}\ \bibnamefont {Fanto}}, \bibinfo {author}
  {\bibfnamefont {P.~M.}\ \bibnamefont {Thomas}}, \bibinfo {author}
  {\bibfnamefont {S.~F.}\ \bibnamefont {Preble}}, \bibinfo {author}
  {\bibfnamefont {A.~M.}\ \bibnamefont {Smith}}, \bibinfo {author}
  {\bibfnamefont {P.~M.}\ \bibnamefont {Alsing}}, \bibinfo {author}
  {\bibfnamefont {M.}~\bibnamefont {Liscidini}},\ and\ \bibinfo {author}
  {\bibfnamefont {J.~E.}\ \bibnamefont {Sipe}},\ }\bibfield  {title} {\bibinfo
  {title} {Truly unentangled photon pairs without spectral filtering},\ }\href
  {https://doi.org/10.1364/OL.42.003638} {\bibfield  {journal} {\bibinfo
  {journal} {Optics Letters}\ }\textbf {\bibinfo {volume} {42}},\ \bibinfo
  {pages} {3638} (\bibinfo {year} {2017})}\BibitemShut {NoStop}%
\bibitem [{\citenamefont {Legero}\ \emph {et~al.}(2004)\citenamefont {Legero},
  \citenamefont {Wilk}, \citenamefont {Hennrich}, \citenamefont {Rempe},\ and\
  \citenamefont {Kuhn}}]{legero2004b}%
  \BibitemOpen
  \bibfield  {author} {\bibinfo {author} {\bibfnamefont {T.}~\bibnamefont
  {Legero}}, \bibinfo {author} {\bibfnamefont {T.}~\bibnamefont {Wilk}},
  \bibinfo {author} {\bibfnamefont {M.}~\bibnamefont {Hennrich}}, \bibinfo
  {author} {\bibfnamefont {G.}~\bibnamefont {Rempe}},\ and\ \bibinfo {author}
  {\bibfnamefont {A.}~\bibnamefont {Kuhn}},\ }\bibfield  {title} {\bibinfo
  {title} {Quantum {{Beat}} of {{Two Single Photons}}},\ }\href
  {https://doi.org/10.1103/PhysRevLett.93.070503} {\bibfield  {journal}
  {\bibinfo  {journal} {Physical Review Letters}\ }\textbf {\bibinfo {volume}
  {93}},\ \bibinfo {pages} {070503} (\bibinfo {year} {2004})}\BibitemShut
  {NoStop}%
\bibitem [{\citenamefont {Legero}\ \emph {et~al.}(2003)\citenamefont {Legero},
  \citenamefont {Wilk}, \citenamefont {Kuhn},\ and\ \citenamefont
  {Rempe}}]{legero2003}%
  \BibitemOpen
  \bibfield  {author} {\bibinfo {author} {\bibfnamefont {T.}~\bibnamefont
  {Legero}}, \bibinfo {author} {\bibfnamefont {T.}~\bibnamefont {Wilk}},
  \bibinfo {author} {\bibfnamefont {A.}~\bibnamefont {Kuhn}},\ and\ \bibinfo
  {author} {\bibfnamefont {G.}~\bibnamefont {Rempe}},\ }\bibfield  {title}
  {\bibinfo {title} {Time-resolved two-photon quantum interference},\ }\href
  {https://doi.org/10.1007/s00340-003-1337-x} {\bibfield  {journal} {\bibinfo
  {journal} {Applied Physics B}\ }\textbf {\bibinfo {volume} {77}},\ \bibinfo
  {pages} {797} (\bibinfo {year} {2003})}\BibitemShut {NoStop}%
\bibitem [{\citenamefont {Christ}\ and\ \citenamefont
  {Silberhorn}(2012)}]{christ2012}%
  \BibitemOpen
  \bibfield  {author} {\bibinfo {author} {\bibfnamefont {A.}~\bibnamefont
  {Christ}}\ and\ \bibinfo {author} {\bibfnamefont {C.}~\bibnamefont
  {Silberhorn}},\ }\bibfield  {title} {\bibinfo {title} {Limits on the
  deterministic creation of pure single-photon states using parametric
  down-conversion},\ }\href {https://doi.org/10.1103/PhysRevA.85.023829}
  {\bibfield  {journal} {\bibinfo  {journal} {Physical Review A}\ }\textbf
  {\bibinfo {volume} {85}},\ \bibinfo {pages} {023829} (\bibinfo {year}
  {2012})}\BibitemShut {NoStop}%
\bibitem [{\citenamefont {Bonneau}\ \emph {et~al.}(2015)\citenamefont
  {Bonneau}, \citenamefont {Mendoza}, \citenamefont {O'Brien},\ and\
  \citenamefont {Thompson}}]{bonneau2015}%
  \BibitemOpen
  \bibfield  {author} {\bibinfo {author} {\bibfnamefont {D.}~\bibnamefont
  {Bonneau}}, \bibinfo {author} {\bibfnamefont {G.~J.}\ \bibnamefont
  {Mendoza}}, \bibinfo {author} {\bibfnamefont {J.~L.}\ \bibnamefont
  {O'Brien}},\ and\ \bibinfo {author} {\bibfnamefont {M.~G.}\ \bibnamefont
  {Thompson}},\ }\bibfield  {title} {\bibinfo {title} {Effect of loss on
  multiplexed single-photon sources},\ }\href
  {https://doi.org/10.1088/1367-2630/17/4/043057} {\bibfield  {journal}
  {\bibinfo  {journal} {New Journal of Physics}\ }\textbf {\bibinfo {volume}
  {17}},\ \bibinfo {pages} {043057} (\bibinfo {year} {2015})}\BibitemShut
  {NoStop}%
\bibitem [{\citenamefont {Roztocki}\ \emph {et~al.}(2017)\citenamefont
  {Roztocki}, \citenamefont {Kues}, \citenamefont {Reimer}, \citenamefont
  {Wetzel}, \citenamefont {Little}, \citenamefont {Chu}, \citenamefont {Moss},\
  and\ \citenamefont {Morandotti}}]{roztocki2017}%
  \BibitemOpen
  \bibfield  {author} {\bibinfo {author} {\bibfnamefont {P.}~\bibnamefont
  {Roztocki}}, \bibinfo {author} {\bibfnamefont {M.}~\bibnamefont {Kues}},
  \bibinfo {author} {\bibfnamefont {C.}~\bibnamefont {Reimer}}, \bibinfo
  {author} {\bibfnamefont {B.}~\bibnamefont {Wetzel}}, \bibinfo {author}
  {\bibfnamefont {B.~E.}\ \bibnamefont {Little}}, \bibinfo {author}
  {\bibfnamefont {S.~T.}\ \bibnamefont {Chu}}, \bibinfo {author} {\bibfnamefont
  {D.~J.}\ \bibnamefont {Moss}},\ and\ \bibinfo {author} {\bibfnamefont
  {R.}~\bibnamefont {Morandotti}},\ }\bibfield  {title} {\bibinfo {title}
  {Pulsed quantum frequency combs from an actively mode-locked intra-cavity
  generation scheme},\ }in\ \href@noop {} {\emph {\bibinfo {booktitle} {2017
  {{Conference}} on {{Lasers}} and {{Electro}}-{{Optics}} ({{CLEO}})}}}\
  (\bibinfo {year} {2017})\ pp.\ \bibinfo {pages} {1--1}\BibitemShut {NoStop}%
\bibitem [{\citenamefont {Agha}\ \emph {et~al.}(2012)\citenamefont {Agha},
  \citenamefont {Davan{\c c}o}, \citenamefont {Thurston},\ and\ \citenamefont
  {Srinivasan}}]{agha2012}%
  \BibitemOpen
  \bibfield  {author} {\bibinfo {author} {\bibfnamefont {I.}~\bibnamefont
  {Agha}}, \bibinfo {author} {\bibfnamefont {M.}~\bibnamefont {Davan{\c c}o}},
  \bibinfo {author} {\bibfnamefont {B.}~\bibnamefont {Thurston}},\ and\
  \bibinfo {author} {\bibfnamefont {K.}~\bibnamefont {Srinivasan}},\ }\bibfield
   {title} {\bibinfo {title} {Low-noise chip-based frequency conversion by
  four-wave-mixing {{Bragg}} scattering in {{SiN}}{\textsubscript{x}}
  waveguides},\ }\href {https://doi.org/10.1364/OL.37.002997} {\bibfield
  {journal} {\bibinfo  {journal} {Optics Letters}\ }\textbf {\bibinfo {volume}
  {37}},\ \bibinfo {pages} {2997} (\bibinfo {year} {2012})}\BibitemShut
  {NoStop}%
\bibitem [{\citenamefont {Li}\ \emph {et~al.}(2016)\citenamefont {Li},
  \citenamefont {Davan{\c c}o},\ and\ \citenamefont {Srinivasan}}]{li2016a}%
  \BibitemOpen
  \bibfield  {author} {\bibinfo {author} {\bibfnamefont {Q.}~\bibnamefont
  {Li}}, \bibinfo {author} {\bibfnamefont {M.}~\bibnamefont {Davan{\c c}o}},\
  and\ \bibinfo {author} {\bibfnamefont {K.}~\bibnamefont {Srinivasan}},\
  }\bibfield  {title} {\bibinfo {title} {Efficient and low-noise
  single-photon-level frequency conversion interfaces using silicon
  nanophotonics},\ }\href {https://doi.org/10.1038/nphoton.2016.64} {\bibfield
  {journal} {\bibinfo  {journal} {Nature Photonics}\ }\textbf {\bibinfo
  {volume} {10}},\ \bibinfo {pages} {406} (\bibinfo {year} {2016})}\BibitemShut
  {NoStop}%
\bibitem [{\citenamefont {Li}\ \emph {et~al.}(2017)\citenamefont {Li},
  \citenamefont {Sun},\ and\ \citenamefont {Foster}}]{li2017}%
  \BibitemOpen
  \bibfield  {author} {\bibinfo {author} {\bibfnamefont {K.}~\bibnamefont
  {Li}}, \bibinfo {author} {\bibfnamefont {H.}~\bibnamefont {Sun}},\ and\
  \bibinfo {author} {\bibfnamefont {A.~C.}\ \bibnamefont {Foster}},\ }\bibfield
   {title} {\bibinfo {title} {Four-wave mixing {{Bragg}} scattering in
  hydrogenated amorphous silicon waveguides},\ }\href
  {https://doi.org/10.1364/OL.42.001488} {\bibfield  {journal} {\bibinfo
  {journal} {Optics Letters}\ }\textbf {\bibinfo {volume} {42}},\ \bibinfo
  {pages} {1488} (\bibinfo {year} {2017})}\BibitemShut {NoStop}%
\bibitem [{\citenamefont {Bell}\ \emph {et~al.}(2017)\citenamefont {Bell},
  \citenamefont {Xiong}, \citenamefont {Marpaung}, \citenamefont {McKinstrie},\
  and\ \citenamefont {Eggleton}}]{bell2017}%
  \BibitemOpen
  \bibfield  {author} {\bibinfo {author} {\bibfnamefont {B.~A.}\ \bibnamefont
  {Bell}}, \bibinfo {author} {\bibfnamefont {C.}~\bibnamefont {Xiong}},
  \bibinfo {author} {\bibfnamefont {D.}~\bibnamefont {Marpaung}}, \bibinfo
  {author} {\bibfnamefont {C.~J.}\ \bibnamefont {McKinstrie}},\ and\ \bibinfo
  {author} {\bibfnamefont {B.~J.}\ \bibnamefont {Eggleton}},\ }\bibfield
  {title} {\bibinfo {title} {Uni-directional wavelength conversion in silicon
  using four-wave mixing driven by cross-polarized pumps},\ }\href
  {https://doi.org/10.1364/OL.42.001668} {\bibfield  {journal} {\bibinfo
  {journal} {Optics Letters}\ }\textbf {\bibinfo {volume} {42}},\ \bibinfo
  {pages} {1668} (\bibinfo {year} {2017})}\BibitemShut {NoStop}%
\end{thebibliography}
\end{document}